\documentclass[%
 reprint,
 amsmath,amssymb,
 aps,
prx,
longbibliography
]{revtex4-2}

\usepackage{graphicx}
\usepackage{dcolumn}
\usepackage{bm}

\usepackage{textcomp}
\usepackage{xfrac}
\usepackage{tikz,graphicx}
\usepackage{circuitikz}
\usepackage{dcolumn}
\usepackage{epstopdf} 
\usepackage{times}
\usepackage{color}
\usepackage{dsfont}
\usepackage[normalem]{ulem}
\usepackage{mathrsfs}

\newcommand{\Tr}{\operatorname{Tr}}

\renewcommand{\H}{\mathcal{H}}

\newcommand{\totimes}{\vcenter{\hbox{$\scriptstyle\otimes$}}}
\newcommand{\sigmaz}{\sigma_{\text{\tiny Z}}}
\newcommand{\sigmax}{\sigma_{\text{\tiny X}}}
\newcommand{\sigmay}{\sigma_{\text{\tiny Y}}}
\newcommand{\e}{\text{e}}
\newcommand{\tsub}[1]{$_{\text{#1}}$}

\begin{document}


\title{Sensitivity of quantum information to environment perturbations \\measured with a non-local out-of-time-order correlation function}

\author{Mohamad Niknam}
\email{mniknam@uwaterloo.ca}
\affiliation{Institute for Quantum Computing, University of Waterloo, Waterloo, ON, Canada, N2L3G1}
\affiliation{Department of Physics, University of Waterloo, Waterloo, ON, Canada, N2L3G1}
\author{Lea F. Santos}
\affiliation{Department of Physics, Yeshiva University, New York, New York 10016, USA}
\author{ and David G. Cory}
\affiliation{Institute for Quantum Computing, University of Waterloo, Waterloo, ON, Canada, N2L3G1}
\affiliation{Department of Chemistry, University of Waterloo, Waterloo, ON, Canada, N2L3G1}
\affiliation{Perimeter Institute, University of Waterloo, Waterloo, Canada, N2L2Y5}
\affiliation{Canadian Institute For Advanced Research, Toronto, ON, Canada, M5G1Z8}

\date{\today}

\begin{abstract}
In a quantum system coupled with a non-Markovian environment, quantum information may flow out of or in to the system.   Measuring quantum information flow and its sensitivity to perturbations is important for a better understanding of open quantum systems and for the implementation of quantum technologies. Information gets shared between a quantum system and its environment by means of system-environment correlations (SECs) that grow during their interaction.  We design a nuclear magnetic resonance (NMR) experiment to directly observe the evolution of the SECs and use the second moment of their distribution as a natural metric for quantifying the flow of information. In a second experiment, by accounting for the environment dynamics, we study the sensitivity of the shared quantum information to perturbations in the environment. The metric used in this case is a non-local out-of-time-order correlation function (OTOC).  By analyzing the decay of the OTOC as a function of the SEC spread, instead of the evolution time, we are able to demonstrate its exponential behavior.
\end{abstract}

\pacs{Valid PACS appear here}
\maketitle


\section{\label{Introduction}Introduction}

The development of quantum technologies is obstructed by the loss of quantum properties caused by interactions with the environment that lead to decoherence~\cite{Bennett2000,Myatt2000,Khaetskii02,Coish2004,Cucchietti2005}.
Quantum information is shared with the environment by means of system-environment correlations (SECs)~\cite{Yang2008,Pernice2012,Ma2014}. In the case of interactions with a non-Markovian environment, the SECs may lead to the flow of quantum information back to the system~\cite{Breuer2009,Breuer2016}.  
Interferences arising from this backflow may be catastrophic to quantum information processes taking place in the system. 
The purpose of this work is to investigate the growth of the SECs as a function of time and to gauge how
susceptible they are to perturbations in the environment.  
An open question with respect to the  latter point is whether environment perturbations can be used to reduce information backflow.

Our experiment is well-equipped to directly measure correlations between the system and the environment. It builds upon solid-state nuclear magnetic resonance (NMR) methods that have been employed to detect multiple-quantum coherences in homonuclear many-body systems~\cite{Baum1985,Munowitz1987,Lacelle1993,Ramanathan2003,Krojanski2004,Cho2005,vanBeek2005,Lovric2011,Sanchez2016}. These methods have recently been used for the investigation of multiple-quantum coherences in ion traps~\cite{Garttner2017,Garttner2018}.
A change in the encoding basis has allowed for the observation of multi-spin dynamics of correlation growth during the free induction decay experiment~\cite{Ramanathan2003,Cho2005}. Here,
we extend these methods so that they can be used for composite heteronuclear systems to measure the growth of correlations with the environment. 

We consider a central spin model, which consists of a single spin-1/2 interacting with environment spins of another spin species that may also be coupled~\cite{DeGennes58,Anderson1959,Gaudin76,Stamp2000}.  Quantum information initially resides in the central spin and is later shared with the environment in the form of multi-spin SECs. NMR techniques make it possible to separate the system-environment evolution from the internal evolution of the environment spins~\cite{Rhim73_mrev8,Rhim74_tuneupcycles,HaeberlenBook,DybowskiBook}. This allows us to examine the impact of each process individually.

We design two different NMR echo experiments. In the first, the environment dynamics is off. We analyze the evolution of the SECs in time and discuss how the second moment of the distribution of these correlations can be used to quantify the flow of quantum information between the system and the environment. In analogy with recent studies~\cite{Garttner2018}, this metric  is related to the quantum Fisher information.

In the second experiment, the environment dynamics is turned on, causing the scrambling of quantum information. To quantify the sensitivity of information to scrambling, we employ a non-local out-of-time-order correlation function (OTOC).  Despite great theoretical interest, very few experiments have had access to this quantity. 
Since NMR echo techniques can reverse time evolution, the OTOC function may be used to study the decay of echo signal amplitude. Experimental implementations of the OTOC was previously studied in closed systems of ion traps~\cite{Garttner2017,Garttner2018} and nuclear spins~\cite{Li2017,Wei2018}.

The OTOC has become a prominent quantity in the analysis of the scrambling of quantum information in black holes and many-body quantum systems. It has been conjectured that the exponential behavior of this quantity should be an indicator of quantum chaos~\cite{Kitaev2015,Maldacena2016}, the exponential rate being associated with the classical Lyapunov exponent. This correspondence has so far been confirmed theoretically for one-body chaotic systems~\cite{Rozenbaum2017,Rozenbaum18} and for the Dicke model~\cite{Chavez-Carlos:2018}, but not yet experimentally. In the case of many-body quantum systems, existing NMR experiments have used the Loschmidt echo and shown that depending on 
the interaction Hamiltonian, both Gaussian and exponential decays can be observed~\cite{Usaj-Levstein1998,Levstein-Pastawski1998}.

In the second echo experiment of this work, we employ an OTOC function that involves 
one global operator, to investigate the sensitivity of the echo signal to the scrambling of SECs in the environment.  For the central spin system, the decay of this non-local OTOC function with respect to the evolution time is Gaussian. However, an exponential behavior is revealed when the non-local OTOC is studied as a function of the spread of the correlations between system and environment (that is, as a function of the Hamming weight spread of SECs). The capability of our experiments to directly measure these correlations is an essential ingredient for describing  the flow of quantum information and for uncovering the exponential decay of the non-local OTOC.

\section{\label{MCD}Mapping the system-environment correlations}

\subsection{\label{sampledisc}Sample description}
\begin{figure}[!bt]
	\begin{center}
		\includegraphics[scale=0.35]{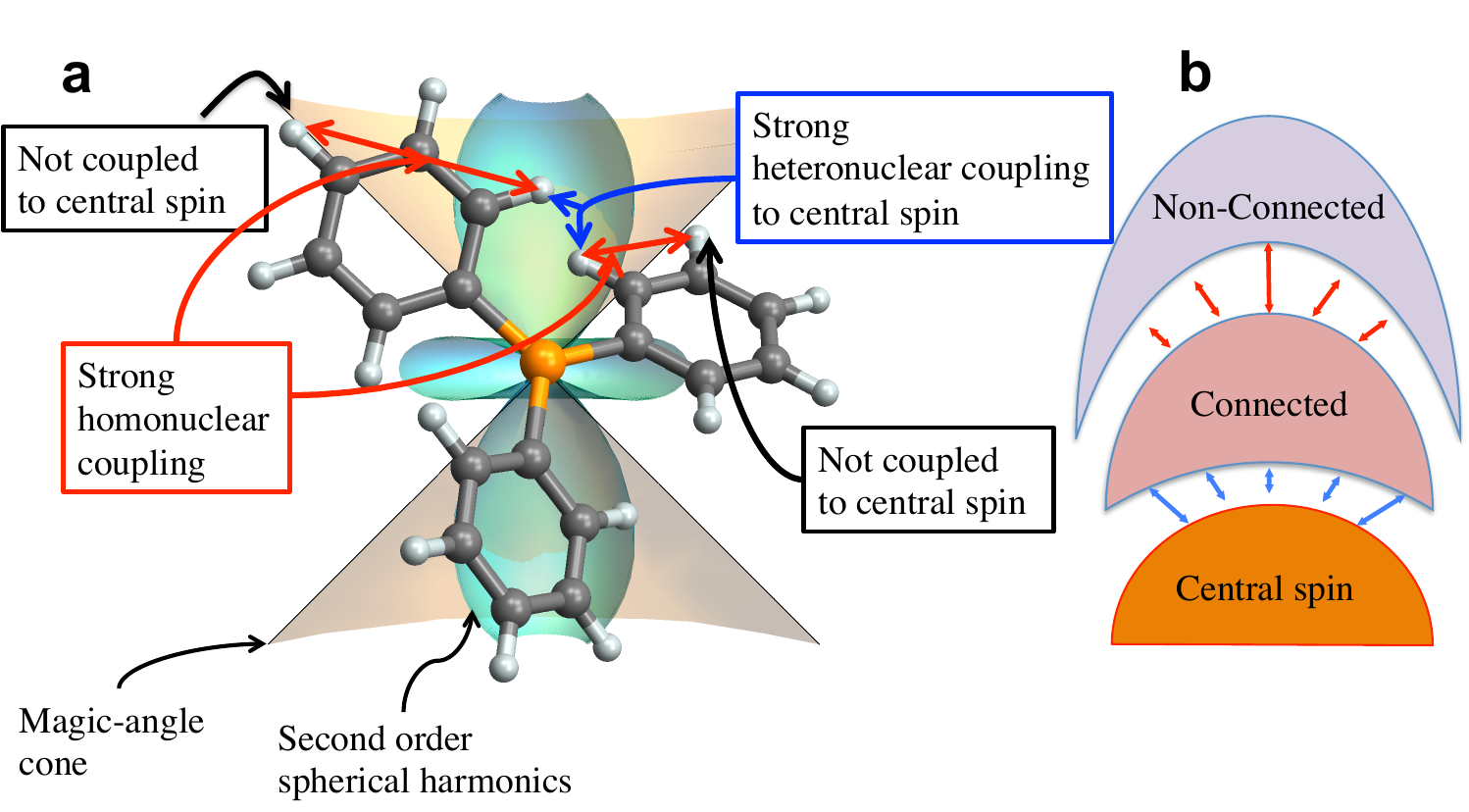} 
	\end{center}	  	 	 	 	
	\caption{ Sample structure matching the central spin model. The Triphenylphosphine molecule shown in panel (a) has a \textsuperscript{31}$ \mathrm{P} $ nucleus at the central spin position and fifteen \textsuperscript{1}$ \mathrm{H} $ nuclei as the environment spins. The NMR experiment is performed on an ensemble of these molecules in random orientations.  Due to the angular dependence of the dipolar interaction, environment spins located near the two magic-angle cones (shaded area) are very weakly coupled to the central spin. Therefore, environment spins may be divided into a connected and a non-connected group, as sketched in panel  (b).  
	}
	\label{dipolarstructure}
\end{figure}

The sample studied is a polycrystalline solid at room temperature composed of an ensemble of Triphenylphosphine molecules, as shown in Fig.~\ref{dipolarstructure}~(a) (see also Appendix~\ref{sample}). Each molecule has a central \textsuperscript{31}P nuclear spin coupled to fifteen \textsuperscript{1}H environment spins via the heteronuclear dipolar interaction
\begin{equation} \label{hetHam}
\H_{SE} =   \sum_{j}^{\text{N}} \omega_j  \sigmaz^{\text{cs}} \totimes \sigmaz^j \totimes \mathds{1}^{\tiny\totimes N-1} ,
\end{equation}
where `cs' stands for central spin, $N=15$ is the number of environment spins, and $\sigma_{\text{\tiny X,Y, Z}}^j$ represent Pauli matrices for the  $ j^{\text{th}} $ spin. The dipolar Hamiltonian is a second rank spherical tensor, where the coupling constant $\omega_j \propto (3 \cos^2 \theta_j -1)/r_j^3 $ has radial and angular dependence on the vector  $ \vec{r}_j $ connecting the central spin to the $j^{\text{th}}$ spin in the environment. $\theta_j$ is the angle between $\vec{r}_j$ and the static field of the NMR magnet, which is along $ z $. These coupling constants for our sample are typically lower than 8  kHz  (see Appendix~\ref{sample}). An environment spin located on the cone defined by the magic angle $ \theta_{\text{\tiny M}} $, with $3 \cos^2 \theta_{\text{\tiny M}}-1=0 $, does not interact with the central spin. The orientation of the sample molecule illustrated in Fig.~\ref{dipolarstructure}~(a) is such that two of the environment spins lie on the magic-angle cone and consequently, do not interact with the central spin. They belong to the ``non-connected'' group, while environment spins coupled to the central spin are part of the ``connected" group, as sketched in Fig.~\ref{dipolarstructure}~(b). The size of the connected group grows in time (see Appendix~\ref{clustersizegrowth}).

The central spin is initially in the state $\rho^{\text{cs}} (0)=[\mathds{1}+\epsilon\sigma_{\text{\tiny X}}]/2 $, where $ \epsilon $ is the strength of the nuclear spin polarization which is of the order of $ 10^{-5} $ at room temperature. In what follows, we drop the identity operator to simplify the notation, since it does not lead to any observable signal.
The $ N $ spins in the environment are initially in the maximally mixed state $\rho^{\text{E}} (0)=(\mathds{1}/2)^{\totimes N} $, with no correlations. Thus, the initial state of the composite system is uncorrelated, $ \rho(0)=\rho^{\text{cs}} (0)\totimes(\mathds{1}/2)^{\totimes N} $.

\subsection{\label{MCDexp}Correlation detection experiment}
\begin{figure}[!hbt]
	\includegraphics*[scale=0.4]{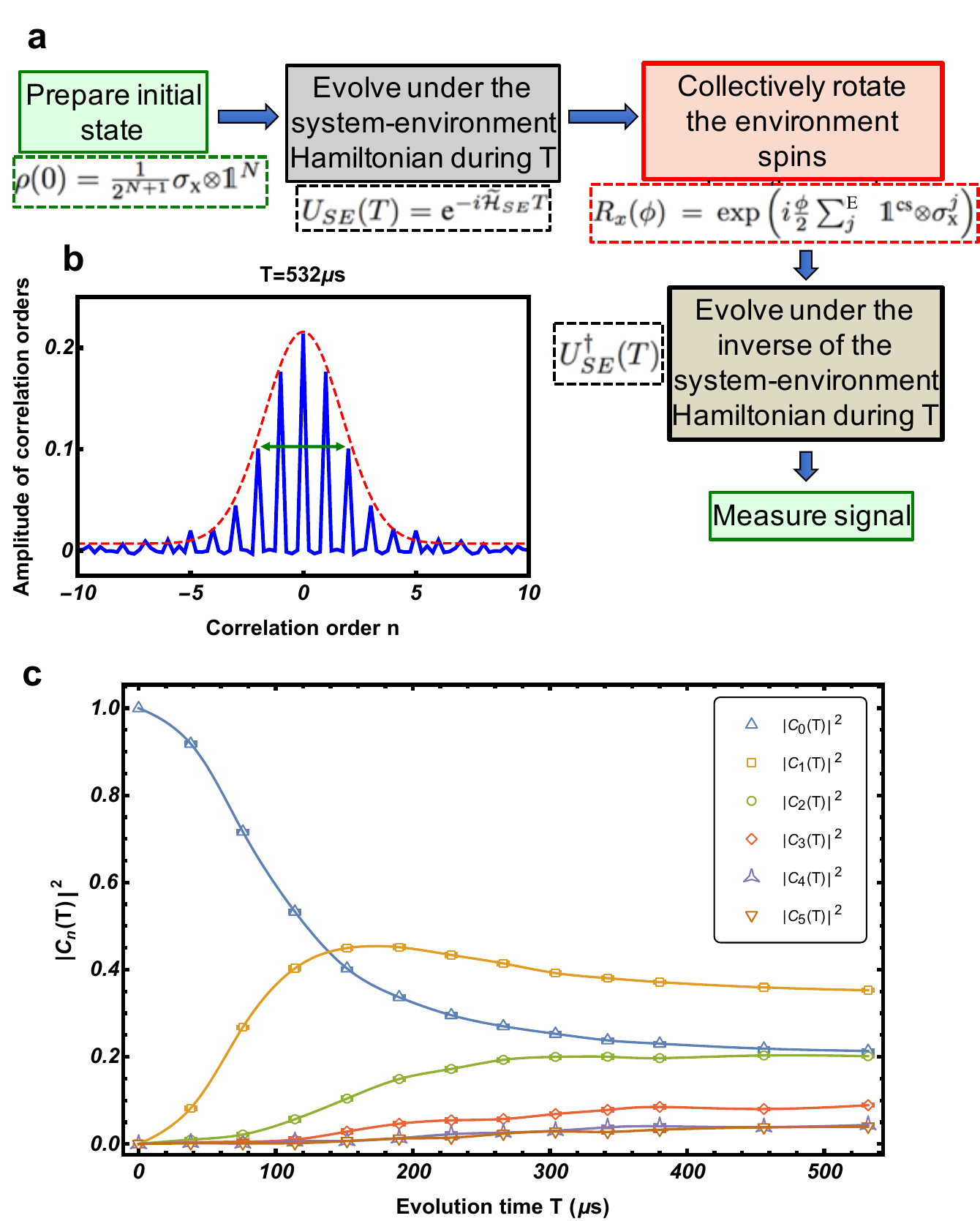}	  	 	 	 	
	\caption{The multi-spin correlation detection (MCD) experiment.The stages of the MCD experiment are sketched in panel (a). The plot for the amplitude of each correlation order in panel (b) displays the distribution of the system-environment correlations (SECs) at a chosen time $T$. The dashed line is a Gaussian fit and the arrow indicates the Hamming weight spread. The sum of the amplitudes for $ +n $ and $ -n $ for each order gives $ |C_n(T)|^2 $. The evolution of these correlation amplitudes in time is plotted in panel (c). The error bars, corresponding to the inverse of the signal-to-noise ratio,  are very small and not visible in panel {\bf c}.  }
	\label{Fig:MCD}
\end{figure}

We design an echo experiment, which we call multi-spin correlation detection (MCD), to measure the correlation growth between the central spin and the environment spins. 
The stages of the MCD experiment are sketched in Fig.~\ref{Fig:MCD}~(a). During the evolution time $T$, the environment self-interaction is averaged to zero using the MREV-8 pulse sequence~\cite{Rhim73_mrev8}, while the remaining system-environment interaction correlates the two. The dynamics in the composite Hilbert space is described by the unitary propagator $U_{SE}(T) = \e^{-i\widetilde{\H}_{SE}T} $, where $ \widetilde{\H}_{SE} $ represents the system-environment interaction Hamiltonian in the toggling frame of the MREV-8 pulse sequence~\cite{HaeberlenBook} (see Appendix~\ref{NMRpulse}). 

After the evolution with the system-environment Hamiltonian, the resulting density matrix is 
\begin{eqnarray}\label{powexp}
\rho(T) \! &\!=\!&  \! U_{SE}(T) \rho(0) U_{SE}^{\dagger}(T) \\
\! &\!=\!& \! \rho(0) +i T [\rho(0) ,\widetilde{\H}_{SE}] - \frac{T^2}{2} [ [\rho(0) ,\widetilde{\H}_{SE}],\widetilde{\H}_{SE}]
+ \ldots .
\nonumber 
\end{eqnarray}
This equation indicates that at short times only the environment spins strongly interacting with the central spin affect the dynamics. The effects of the nested commutators, which are associated with the multi-spin SECs involving weaker interacting spins, become more pronounced as time evolves. Therefore, the evolution of the composite system can be equivalently described using the number of coupled spins and the weight $\mathscr{C}_n(T)$ of each cluster as follows (see Appendix~\ref{numberofcorrspins}),
\begin{eqnarray} 
\rho{(T)}&=& \mathscr{C}_0(T) \sigmax^{\text{cs}} \totimes \mathds{1}^{\tiny\totimes N}  \label{eq:rhotsigmaz}\\
&+& \mathscr{C}_1(T) \sum_{j}^{N}\sigmay^{\text{cs}}  \totimes \sigmaz^j \totimes \mathds{1}^{\tiny\totimes N-1} \nonumber\\
&+& \mathscr{C}_2(T) \sum_{j\neq k}^{N}\sigmax^{\text{cs}}  \totimes \sigmaz^j \totimes \sigmaz^k \totimes \mathds{1}^{\tiny\totimes N-2} \nonumber\\
&+& \cdots  . \nonumber
\end{eqnarray}
The observable signal $ S(T) $ from the central spin corresponds to the inner product of the reduced state of the central spin and the measurement operator, $S(T)=\Tr[\Tr_{\text{E}}[\rho(T)].\sigma^{\text{\tiny CS}}_{\text{\tiny X}}]$. Notice that only the first line of Eq.~(\ref{eq:rhotsigmaz}) survives the partial trace. Since $\sum |{\mathscr C}_n(T)|^2 =1$, as the multi-spin correlations increase, the observable signal from the central spin decays. This is the free induction decay.

The measurement operator in the NMR experiment is a transverse single-spin operator, so only the single-spin term in the density matrix induces  NMR signal, while the multi-spin correlated terms are not directly observable.   In our case, the reduced state of the environment would not reveal the evolution of SECs, because $\Tr_{\text{cs}}[\rho(T)] = \mathds{1} $.  In order to  observe the growth of the SECs, we implement the multiple-quantum coherence method for encoding the coherence orders and then detect them through the central spin, which is our probe. In the MCD experiment, by collectively rotating the environment spins along the $x$-axis, $R_x(\phi)=\exp\left( i \frac{\phi}{2} \sum_j \mathds{1}^{\text{cs}} \totimes \mathds{1}^1 \totimes \cdots \totimes \sigma_{\text{\tiny X}}^j\totimes \cdots\totimes \mathds{1}^N  \right)$, we manage to get the  coherence order encoded in a phase factor
\begin{equation} \label{eq:phaseencoding}
\rho_{\phi}(T)=R_x(\phi) \rho(T) R_x^{\dagger}(\phi)=\sum_{n} e^{i n \phi} C_n(T) \rho_{n}^{\text x}  .
\end{equation}
In the above,  $ \rho_{n}^{\text x}  $ indicates the subset of  spin operators in Liouville space with correlation order $ n $ with respect to the $ x $-basis. $C_n(T)$ represents the weight of the multi-spin terms with correlation order $n$. In this basis, the ladder operators are  $\Sigma^{j}_{\pm} =\sigmay^j \pm i \sigmaz^j$, and the correlation order is defined as  the absolute value of the number of $\Sigma_{+}$  minus $\Sigma_{-}$ operators. This number represents the Hamming weight.  Consequently, the relevant description of the density matrix in the $ x $-basis is
\begin{equation}\label{eq:rhoxbasis}
\rho{(T)} =\sum_{n} C_n(T) \rho_{n}^{\text x}  .
\end{equation} 
In contrast with Eq.(\ref{eq:rhotsigmaz}), where the multi-spin terms are categorized by the number of correlated spins, in this equation they are distinguished by their correlation orders (see details in the Appendix~\ref{x-basis}).

After the encoding rotation $ R_x(\phi) $, by applying a $ \pi $-pulse to the central spin, which changes the effective Hamiltonian from $ \H_{SE} $ to $ -\H_{SE} $,  the system-environment Hamiltonian is reversed to create an observable echo at time $ 2T $. The observable NMR signal for each encoding angle $ \phi $  is:  
\begin{equation}\label{Sphi2T}
\hspace{-2.4 mm}S_{\phi}(2T) = \Tr [ \Tr_{\text{E}}[\rho_{\phi} (2T)].\sigma^{\text{\tiny CS}}_{\text{\tiny X}} ]=\Tr [\rho_{\phi} (2T).  \sigma^{\text{\tiny CS}}_{\text{\tiny X}} \totimes \mathds{1}^{\tiny {\otimes N}}] , 
\end{equation}
where
\[
\rho_{\phi}(2T) \!=\! U_{SE}^{\dagger}(T)R_x(\phi)U_{SE}(T)\rho (0) U_{SE}^{\dagger}(T)R_x^{\dagger}(\phi)U_{SE}(T).
\]
The encoding angle is incremented in steps determined by the Nyquist rate, $ \frac{2\pi}{2N} $. The  array of echo amplitudes obtained is  Fourier transformed with respect to $ \phi $, generating correlation amplitudes for each correlation order $ n $. The resulting spectrum of correlation amplitudes provides a snapshot of the SECs at each time $T$. As seen in Fig.~\ref{Fig:MCD}~(b), the distribution of the correlation orders has a Gaussian shape. This form emerges because the signal is an ensemble average over various molecules where each one has a binomial distribution of $C_n(T)$'s (see Appendix~\ref{xquantizationcorr}).

A map of the SEC production is obtained by adding the correlation amplitudes for $ n $ and $ -n $ at each evolution time $T$, as depicted in Fig.~\ref{Fig:MCD}~(c). There, we show the six largest amplitudes of $ |C_n(T)|^2 $. At short times, the first term in $C_0(T)$ coincides with the $\mathscr{C}_0(T)$ term and dominates the dynamics. At longer times,  we observe that the values of the $C_0(T)$ and $C_2(T)$ terms approach each other. This is because the production of the $\mathscr{C}_2(T)$ term is responsible for the onset of both $C_2(T)$ and also the terms $\Sigma^j_{+}\totimes\Sigma^k_{-}$ in $C_0(T)$.

\subsection{\label{QIflow}Flow of quantum information}
\begin{figure}[!tb]
	\centering
	\includegraphics*[scale=0.5]{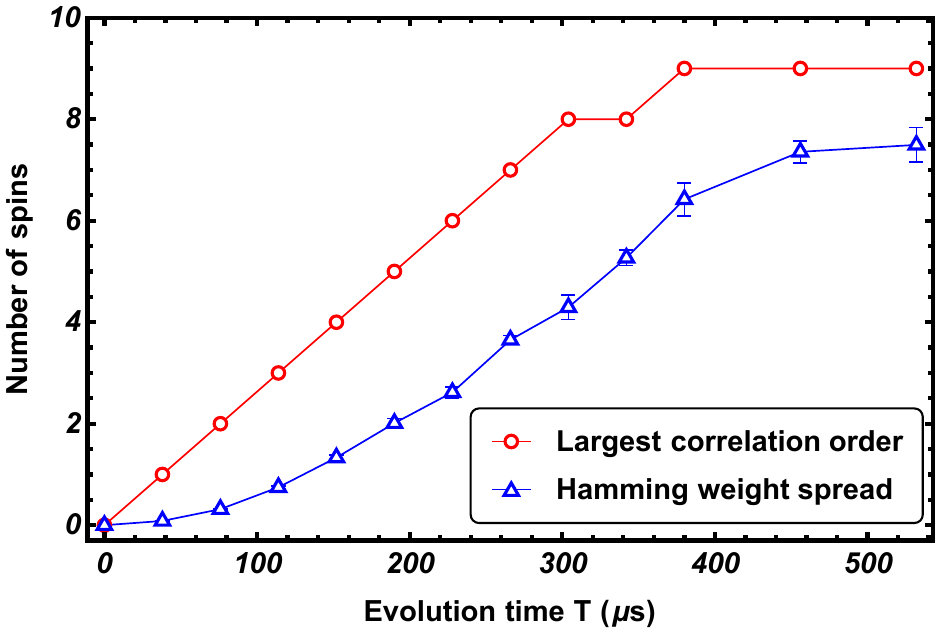}	  	 	 	 	
	\caption{ Hamming weight spread. The second moment of the distribution of multi-spin system-environment correlations (SECs) and the largest correlation order as a function of time are indicators for the growth of the system-environment correlations. The Hamming weight spread is used to quantify the extent of quantum information shared with the environment. 
	}
	\label{Fig:QIflow}
\end{figure}

To quantify the SEC production, one can use the largest correlation order observed in the MCD experiment, 
which is plotted in Fig.~\ref{Fig:QIflow}. However, it becomes more difficult to detect the correlation 
orders as they increase, because the amplitude of the largest correlation order drops exponentially (see Appendix~\ref{xquantizationcorr}).  Alternatively, 
the second moment  of the distribution of the correlation orders,  $\sum_{n} |C_n(T)|^2 n^2$,   is a more reliable experimental measure for quantifying the extent of the correlations.  The second moment (variance) of a binomial distribution centered at 0 is equal to  $ n $. This value is the square of the width of the SEC distribution in  Fig.~\ref{Fig:MCD}~(b).  
Here, we refer to the second moment as the ``Hamming weight spread''. The second moment of the coherence distribution is also used in~\cite{Munowitz1987,Krojanski2004} for quantifying the number of spins involved in the clusters of linked spins in homonuclear solid-state systems. 

In Fig.~\ref{Fig:QIflow}, we show the Hamming weight spread as a function of $ T $, which initially grows slowly and later linearly before saturating. The point of saturation depends on the size of the connected group of environment spins (see Appendix~\ref{clustersizegrowth}). 

In Ref.~\cite{Garttner2018}, it was shown that the second moment of the multiple quantum coherences for many-body systems with mixed states is a lower bound on the quantum Fisher information (QFI). 
In our experiment, the QFI is associated with the information shared between the system and the environment and the rate of its change measures the information flow, as discussed in~\cite{Lu2010}. By extension, the slope of the curve for the Hamming weight spread can be used as a metric to quantify the flow of information between the system and the environment.

\vskip 0.2 cm
\section{\label{Scrambling}Scrambling of information in the environment}

\subsection{\label{MCSDexp}Information scrambling experiment}
In the second experiment, we explore the resistance of the quantum information shared between the system and environment against perturbations in the environment. The latter refer to changes that take place when we turn on the dynamics in the environment. The  homonuclear dipolar Hamiltonian is given by
\begin{equation}
\mathcal{H}_{E} = \mathds{1}^{\text{cs}} \totimes \sum_{j<k}^{N} \Omega_{jk} \left[ \sigma_{\text{\tiny Z}}^j\sigma_{\text{\tiny Z}}^k- \frac{1}{4} (\sigma^j_+ \sigma^k_-+\sigma^j_- \sigma^k_+) \right] ,
\label{Eq:bath}
\end{equation}
where $\Omega_{jk}\propto (3 \cos^2 \theta_{jk} -1)/r_{jk}^3$ is the coupling strength between the environment spins $j$ and $k$, with typical values below 20  kHz  (see Appendix~\ref{sample}). The operators in parenthesis in Eq.~(\ref{Eq:bath}) represent the flip-flop term that swaps the states of pairs of environment spins and scrambles quantum information. The eigenvalues of this Hamiltonian satisfy level statistics given by random matrix theory (see Appendix~\ref{qchaos}), as in quantum systems with chaotic classical counterparts.

\begin{figure}[!b]
	\centering
	\includegraphics*[scale=0.35]{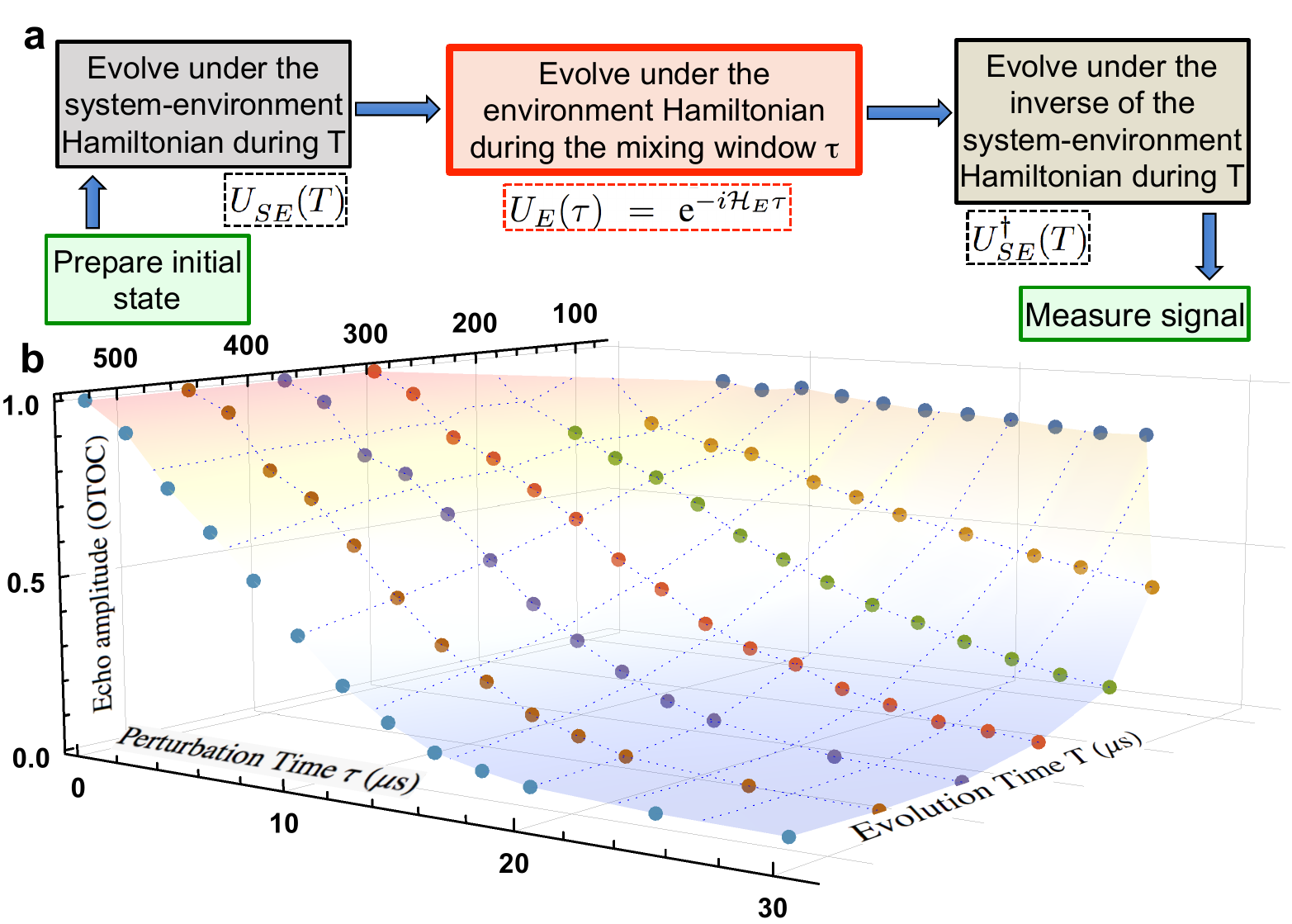}
	\caption{ The OTOC  $F_{\tau} (T)$ measures the sensitivity of quantum information to environment perturbations. Sketch of the steps involved in the OTOC experiment is depicted in (a). The decay of the echo amplitude as a function of the scrambling window length $\tau$ for different values of the evolution time $T$ is shown in (b). For small perturbations (short $\tau$), the central spin for most molecules gets refocused and the OTOC  is nearly independent of $T$, while for large perturbations,  the OTOC  decays significantly with $ T $. The surface  plotted here is a guide for the eye.  }
	\label{Fig:scrambling}
\end{figure}

To analyze the sensitivity of quantum information to scrambling in the environment, we use the echo experiment outlined in Fig.~\ref{Fig:scrambling}~(a). In this experiment, the evolution interval $ T $, where only the coupling between the system and the environment is effective, is followed by a scrambling window of length $\tau$, where only the environment spins interact and the propagator is $U_{E}(\tau)= \e^{-i\H_{E} \tau}$. Information shared with the environment in the course of time $T$ gets scrambled during $\tau$. The length of the scrambling window $\tau$ determines the strength of the environment perturbation. The observable echo signal amplitude at $2T+\tau$ is given by
\begin{equation}
S(2T+\tau) =  \Tr[\rho(2T+\tau). \sigma^{\text{\tiny CS}}_{\text{\tiny X}} \totimes \mathds{1}^{\tiny {\otimes N}}] , \label{Eq:echosignal}
\end{equation}
where
\[
\rho(2 T+\tau) \!=\! U_{SE}^{\dagger}(T) U_E(\tau ) U_{SE}(T) \rho(0) U_{SE}^{\dagger}(T) U_E^{\dagger}(\tau ) U_{SE}(T).
\]
For a fixed value of $ \tau $, as the evolution time $ T $ increases, the overlap between the density matrix $ \rho(2 T+\tau) $ and the initial density matrix decreases, resulting in the decay of the  echo signal.

\subsection{\label{OTOC}Non-local out-of-time-order correlation function}
The signal $S(2T+\tau)$ for a fixed scrambling window $\tau $ can be written in the form of an OTOC function with a non-local operator. The latter is defined as 
\begin{equation}\label{eq:OTOCdef}
F (T) \equiv \left<W^{\dagger} (T)V(0)^{\dagger}W(T)V(0)\right>, 
\end{equation}
where $V(0)$ and $W(T)$ are two unitary operators that commute at $T=0$. We choose $V(0)$ to be proportional to the observable operator for the central spin $V(0)= \sigmax^{\text{cs}}\totimes \mathds{1}^{\otimes N}$ and  consider the environment operator in the Heisenberg picture as the second operator which is non-local in this case
\begin{equation} 
W_{\tau}(T) = U_{SE}^{\dagger}(T) U_E^{\dagger}(\tau ) U_{SE}(T).
\label{Eq:wtauT}
\end{equation}
Therefore, with the assumption of infinite temperature, the expectation value in Eq.~(\ref{eq:OTOCdef}) can be written as
\begin{eqnarray}\label{eq:OTOCTr}
F_{\tau} (T) & \equiv& \Tr[W_{\tau}^{\dagger}(T) V(0) W_{\tau}(T) V(0)] \nonumber\\
&=&2^{N+1}S(2T+\tau) \cdot
\end{eqnarray}

The OTOC function is related to the commutator between $V(0)$ and $W_{\tau}(T)$ as  $\mbox{Re}[F_{\tau} (T)] = \{1- \left<|[W_{\tau} (T) ,V(0)]|^2\right> /2\} $.
As $ T $ becomes larger and $ W_{\tau}(T) $ spreads away from $ W_{\tau}(0) $, the commutator $ [W_{\tau}(T),V(0)] $ increases and the non-local OTOC decreases. Physically, what happens is that as the SECs grow, the environment interaction $ \H_E $ has access to a larger subset of correlated spins and the number of swaps that effectively scramble information increases. Therefore,  the decay of this non-local OTOC quantifies the level of sensitivity of quantum information to perturbations in the environment. 

In Fig.~\ref{Fig:scrambling}~(b), the result of the non-local OTOC decay is presented for various evolution and perturbation times. When $\tau=0$, the state of the central spin is completely refocused (revived) and $F_{\tau=0} (T)=1$ for any  $T$.  This happens because the information that is initially encoded in the central spin is not lost during the evolution time. It is simply stored in the form of multi-spin correlations between the system and the environment spins from the connected group. By reversing the evolution, the information can be recovered in the system.

The refocusing degrades and the echo amplitude decays as $\tau$ increases. During the scrambling window, the flip-flop term of $\H_E$ swaps the states of coupled spin pairs in the environment. As a result, the subsequent evolution under the inverse of the system-environment Hamiltonian can only partially revive the initial state. This situation is aggravated by the existence of the non-connected group of environment spins, which do not develop correlations with the central spin during $T$, but may have their states swapped with those from the connected group during the scrambling window. Information that is shared with the non-connected group cannot be recovered, which ultimately leads to the loss of quantum information in the environment. Consequently, the sensitivity of shared quantum information to perturbations in the environment, depends on the scrambling window length as well as the size of connected and non-connected spin groups in the environment. 

\vskip 0.2 cm
\subsection{\label{expdecay} Effectiveness of information scrambling}
\begin{figure}	 	 
	\includegraphics*[scale=0.5]{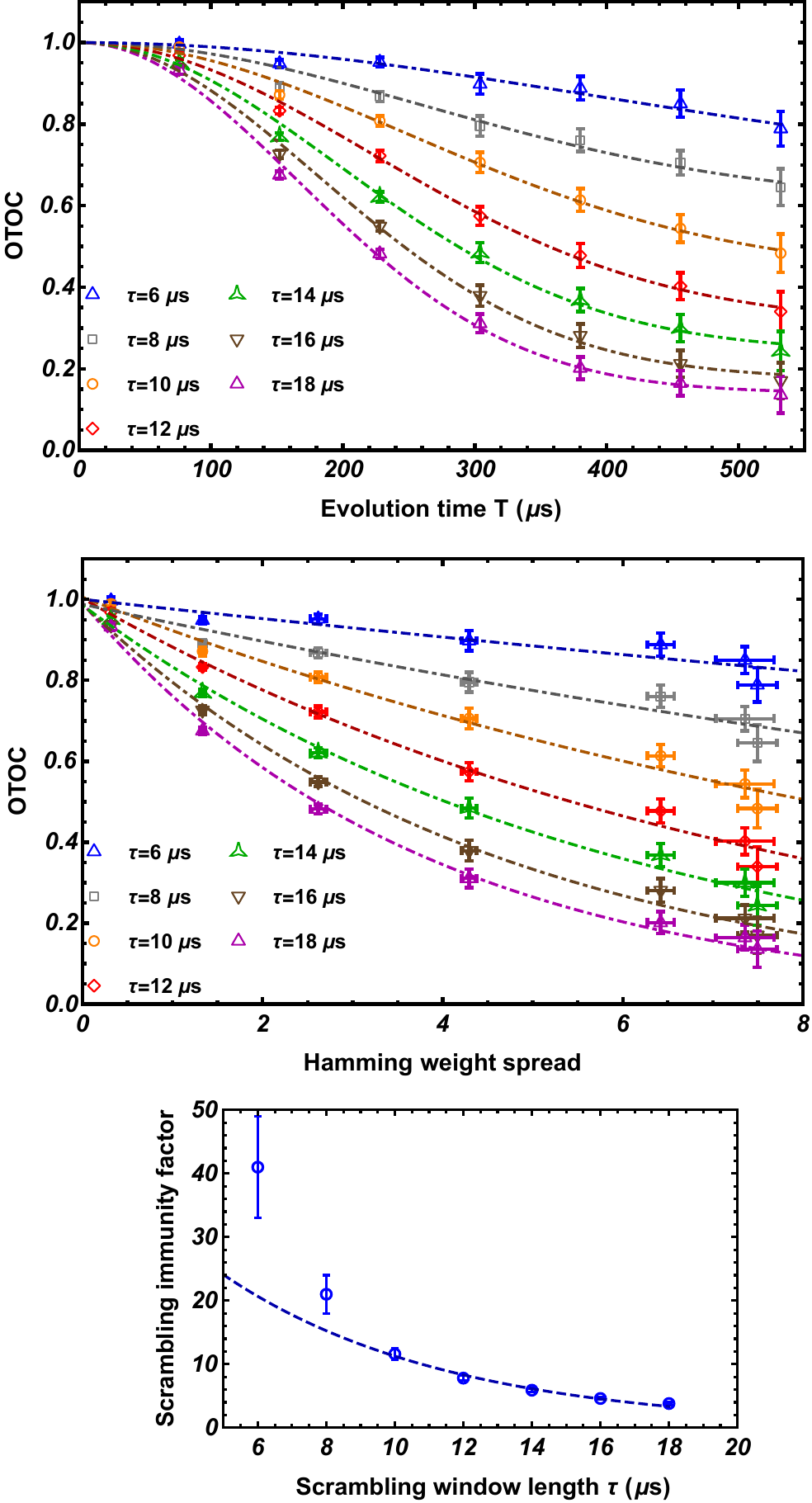}	
	\caption{The non-local OTOC decays exponentially as a function of system-environment correlation spread. Panel (a) depicts the OTOC as a function of the evolution time $T$ and  (b) as a function of the  Hamming weight spread. Each curve in the panels corresponds to a fixed value of the scrambling window $\tau$ (perturbation strength). The data are normalized with respect to the $ \tau=0 $ data set.  The dashed lines in (a) and  (b) are Gaussian and exponential fits, respectively. The exponential behavior is uncovered by analyzing the non-local OTOC decay as a function of the  Hamming weight spread.  Vertical error bars in panel (a) and (b) indicate the inverse of signal-to-noise ratio. Horizontal error bars in panel (b) are those from Fig.~\ref{Fig:QIflow}. In panel (c), the scrambling immunity factor indicates the capability of environment interactions in disrupting the system-environment correlations. This plot indicates that with increasing $ \tau $, even small SECs become sensitive to the environment perturbations.  The data is fitted with an exponential decay curve. The error bars correspond to the errors for the exponential fits in the panel (b). Similar dynamics is observed for swapping coins in a classical coin game (see Appendix~\ref{coingame}). }
	\label{Fig:OTOC}
\end{figure}

For a more detailed analysis of the results for the non-local OTOC presented in Fig.~\ref{Fig:scrambling}~(b), we now show in Fig.~\ref{Fig:OTOC}~(a), the  OTOC as a function of the evolution time $ T $ for different perturbation strengths and compare it with  Fig.~\ref{Fig:OTOC}~(b), where the OTOC is presented as a function of the Hamming weight spread.

As seen in Fig.~\ref{Fig:OTOC}~(a), the OTOC decay in time is well described by a Gaussian function. This is understandable, because the evolution of  environment spins under the homonuclear dipolar interaction $\H_{E}$, examined with a  free induction decay experiment, is known to give a signal decay with Gaussian shape. This behavior is typical of solid-state spin systems~\cite{Abragam,Cho2006}. In the case of our composite system, the scrambling of quantum information that happens only in the environment, is subject to the same homonuclear dipolar Hamiltonian.

The effectiveness of quantum information scrambling, probed with the non-local OTOC decay, increases with the extent of quantum information shared with the environment.  
As seen in Fig.~\ref{Fig:QIflow}, the extent of shared information, which is measured with the Hamming weight spread, does not always grow linearly in time. So in Fig.~\ref{Fig:OTOC}~(b), we shift the perspective by using the Hamming weight spread, obtained in the first experiment, as the variable for the OTOC, instead of time.  The resulting behavior of the OTOC  is exponential, as corroborated by the exponential fits in Fig.~\ref{Fig:OTOC}~(b). 
 One sees that  by removing the non-linear rate of the correlation growth, that is by investigating the OTOC against SEC sizes, the exponential behavior is uncovered. This observation indicates that the amount of quantum information shared between the system and environment determines the  capability of the flip-flop Hamiltonian in scrambling the shared quantum  information. 

We developed a classical coin game  to illustrate the dynamics of spin swaps between connected and non-connected spin groups and to justify the exponential decay of the non-local OTOC (see Appendix~\ref{coingame}). The idea goes as follows. Among $ N $ coins, we randomly flip $ k $. If these same $ k $ coins are flipped a second time, the initial state is recovered. However, if we swap some of the $ N $ coins before the second flip, the final state may be different from the initial one. This happens when some of the $ N-k $ coins get swapped with some of the $ k $ coins. In this game, $ k $ coins represent the connected group of the environment spins, $ N-k $ coins portray the non-connected group, and the number of coin swaps is analogous to the perturbation strength in the environment. We find that similar to Fig.~\ref{Fig:OTOC}~(b), the probability of recovering the initial coin array decreases exponentially as a function of the initial number of flipped coins  $ k $. The inverse rate of this exponential decay characterizes the capability of a fixed number of coin swaps to disrupt the coin array recovery, and is called swap immunity factor. As expected, the swap immunity factor is smaller when a larger number of swaps are performed. This  classical game provides a simplified picture of the mechanism underlying the exponential instability of the composite quantum system.

Motivated by the coin game analysis, we plot in Fig.~\ref{Fig:OTOC}~(c)   the ``scrambling immunity factors'' obtained from the inverse of the decay rates of the exponential fits in Fig.~\ref{Fig:OTOC}~(b). Similar to the concept of the swap immunity factor for a coin array, the  scrambling immunity factor characterizes the capability of the flip-flop Hamiltonian to disrupt the SECs for a given perturbation window $ \tau $, resulting in the incomplete revival of the central spin state. In other words, this factor characterizes the sensitivity of SECs to the scrambling of quantum information in the environment for various perturbation strengths. For the small scrambling windows $ \tau=6,8 $  $\mu s $, the scrambling immunity factor is unreasonably large, as the perturbation is too small and leaves the environment effectively unscrambled for most orientations of the spins. 
For larger perturbation windows,  the scrambling immunity factor decays exponentially with $ \tau $. This shows that, the perturbation strength needed for the effective information scrambling is much smaller for larger SECs.

The scrambling of quantum information is most effective when it involves the non-connected spin group of the environment. Quantum information transferred to the non-connected group can be considered lost. As it cannot produce any echo signal, it does not contribute to the backflow of quantum information to the system. Thus, the scrambling immunity factor provides an upper bound for the effectiveness of environment perturbation in removing quantum information backflow.

We close this section with a discussion about the exponential behavior of OTOCs and the connection with quantum chaos. The search for the quantum counterpart of the exponential instability observed in chaotic classical systems has been a subject of discussion for many years~\cite{Peres1996,Cucchietti2002,Emerson2002,Gorin2006,Elsayed2015,Borgonovi2019,Chirikov1981,SHEPELYANSKY1983,Flambaum2001,Garcia-Mata2018,Luitz2017}. This is now under intense investigation in part due to studies that associate the exponential rate of change of the OTOC with the classical Lyapunov exponent.  So far, most theoretical studies of quantum chaos uses the OTOC to study the overlap of local operators, while here a different implementation with a non-local operator is considered. Whether the observed exponential behavior  does or does not relate to quantum chaos is an interesting open question.   

\vskip 0.2 cm
\section{\label{Conclusion}Conclusion}
We introduced the scrambling immunity factor to characterize the capability of environment perturbations to disrupt the system-environment correlations. Our experiments also enabled us to quantify the flow of quantum information between the system and environment.

We proposed an alternative way to analyze nonequilibrium quantum dynamics, where instead of time, quantities of interest are studied as a function of the SEC spread. This approach uncovered the strong sensitivity of our many-body spin system to environment perturbations, by showing that the non-local OTOC decays exponentially as a function of the Hamming weight spread.

The heart of these experiments is our correlation detection method and this technique is not restricted to the system considered here. It can be used for any quantum system if global control of the environment is available. This is relevant  for many-body quantum systems where the measurements are performed on a subsystem and the rest acts as an environment.

\vskip 0.2 cm
\begin{acknowledgments}
We thank Brian Swingle for useful discussions. The research results communicated here would not be possible without the significant
contributions of the Canada First Research Excellence Fund, Canada Excellence Research Chairs program, Canada Foundation for
Innovation, the Ontario Ministry of Research \& Innovation, Industry Canada and Mike
\& Ophelia Lazaridis. Their support is gratefully acknowledged. L.F.S. was supported by the NSF grant No.~DMR-1603418. 
\end{acknowledgments}

\appendix
\section{\label{sec:experimental}Sample and Methods}

\subsection{\label{sample}Sample}
Triphenylphosphine is a common organophosphorous compound and was obtained from SIGMA-ALDRICH with 99\% purity. To reduce the $ T_1 $ relaxation time of the protons we used  Chromium (III) acetylacetonate as a relaxation agent. $ 1  $ mmol of the sample and $ 0.13 $ mmol of the relaxation agent were resolved in $ 300  $ ml of Chloroform-d and left for crystallization over night. The resulting powder was compressed into a NMR-sphere sample tube which was flame-sealed to best preserve the contents. Using the relaxation agent resulted in the reduction of the proton $ T_1 $ relaxation time from $ 630 \pm30 $s to $ 1.5\pm0.1  $s, as shown in Table~\ref{tab:control}.

\begin{figure}[!h]
	\begin{center}
		\includegraphics[scale=0.5]{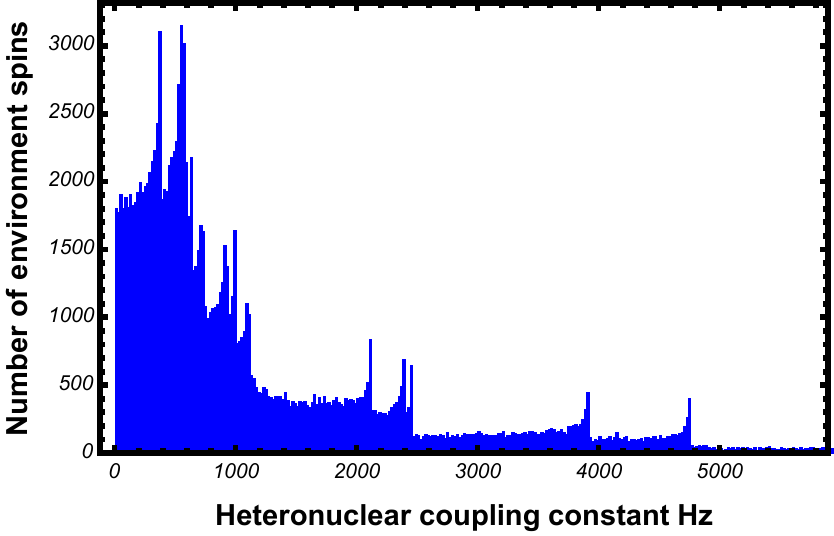} 
	\end{center}	  	 	 	 	
	\caption{A simulation of 10000 random orientations of the Triphenylphosphine molecule is used to  calculate the coupling constants for the heteronuclear dipolar couplings between the central spin and the environment spins given by  Eq.~(\ref{hetHam}). This histogram indicates the distribution of the absolute value of these coupling constants. }
	\label{Fig:whet}
\end{figure}

The distribution of the coupling constants for the system-environment Hamiltonian, Eq.~(\ref{hetHam}), is shown in Fig.~\ref{Fig:whet}. They are calculated for 10000 random orientations of the sample molecule. Notice that the coupling strengths are calculated for static molecules and due to their motion, the observed couplings are slightly weaker in the experiment. These coupling constants determine the system-environment evolution and the consequent map of the correlation amplitudes in Fig.~\ref{Fig:MCD}~(c).

\begin{figure}[!h]
	\begin{center}
		\includegraphics[scale=0.5]{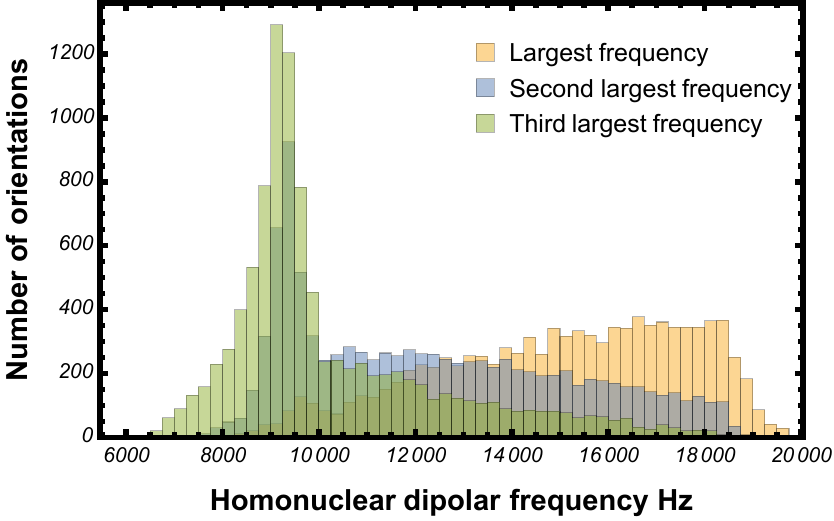} 
	\end{center}	  	 	 	 	
	\caption{This histogram shows the distribution of the absolute values for the three strongest couplings in the environment Hamiltonian, Eq.~(\ref{Eq:bath}). A simulation of 10000 random orientations of the Triphenylphosphine molecule was done to calculate the coupling constants for the homonuclear dipolar interaction between the 15 spins of the environment.}
	\label{Fig:whom}
\end{figure}

A similar simulation is used to study the distribution of the 105 coupling constants for the homonuclear dipolar interaction between the 15 spins of the environment. The scrambling of quantum information in the environment is determined by these coupling constants  as described  by  Eq.~(\ref{Eq:bath}). Figure~\ref{Fig:whom} shows the distribution of the three strongest couplings for each of the 10000 different orientations of the molecule. 

\subsection{\label{clustersizegrowth}Spin groups in the environment}
\begin{figure}[!htb]
	\begin{center}
		\includegraphics[scale=0.5]{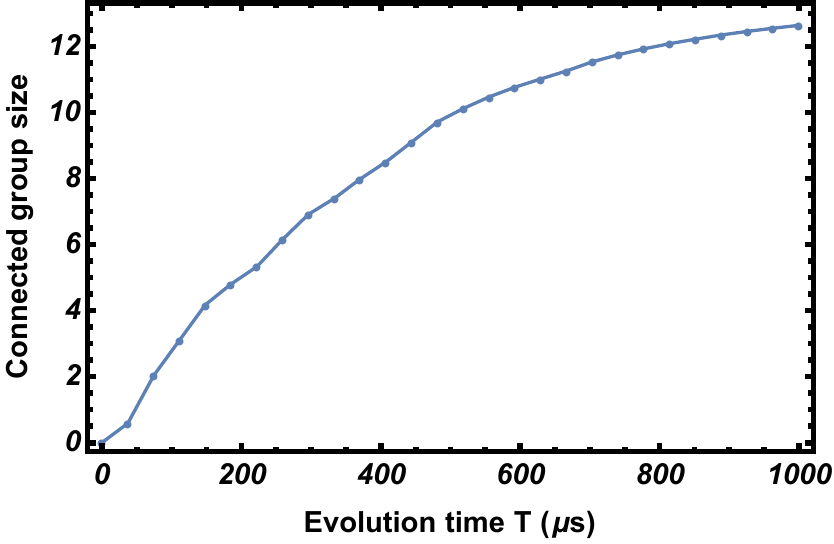} 
	\end{center}	  	 	 	 	
	\caption{A simulation of 10000 random orientations of the Triphenylphosphine molecule is used to estimate the size of the connected spin group for various evolution times in the MCD experiment. }
	\label{Fig:heteronucleardist}
\end{figure}

 The multi-spin correlation growth  and the information scrambling in the environment, both depend on the strength of the dipolar interactions. The dipolar interaction strength depends on the relative orientation of the  spins with respect to the static field of the NMR magnet. The idea of distinguishing connected and non-connected spin groups in the environment can  be explored by considering the number of spins in the environment with a high probability of being correlated to the central spin. The total number of these spins in the environment, increase with the evolution time. Figure~\ref{Fig:heteronucleardist} plots the average number of environment spins that have a probability larger than $ \frac{1}{2} $ for being correlated with the central spin as a function of the evolution time in the MCD experiment.  Heteronuclear dipolar coupling constants are evaluated for 2000 randomly orientated Triphenylphosphine molecules. For the longest evolution time in the MCD experiment,  $ T=532$  $ \mu s $ on average 8.2 environment spins are found to be  more likely to correlate with the central spin.  

\subsection{\label{NMRpulse}NMR experiments}
\begin{figure}
	\begin{center}
		\includegraphics[scale=0.4]{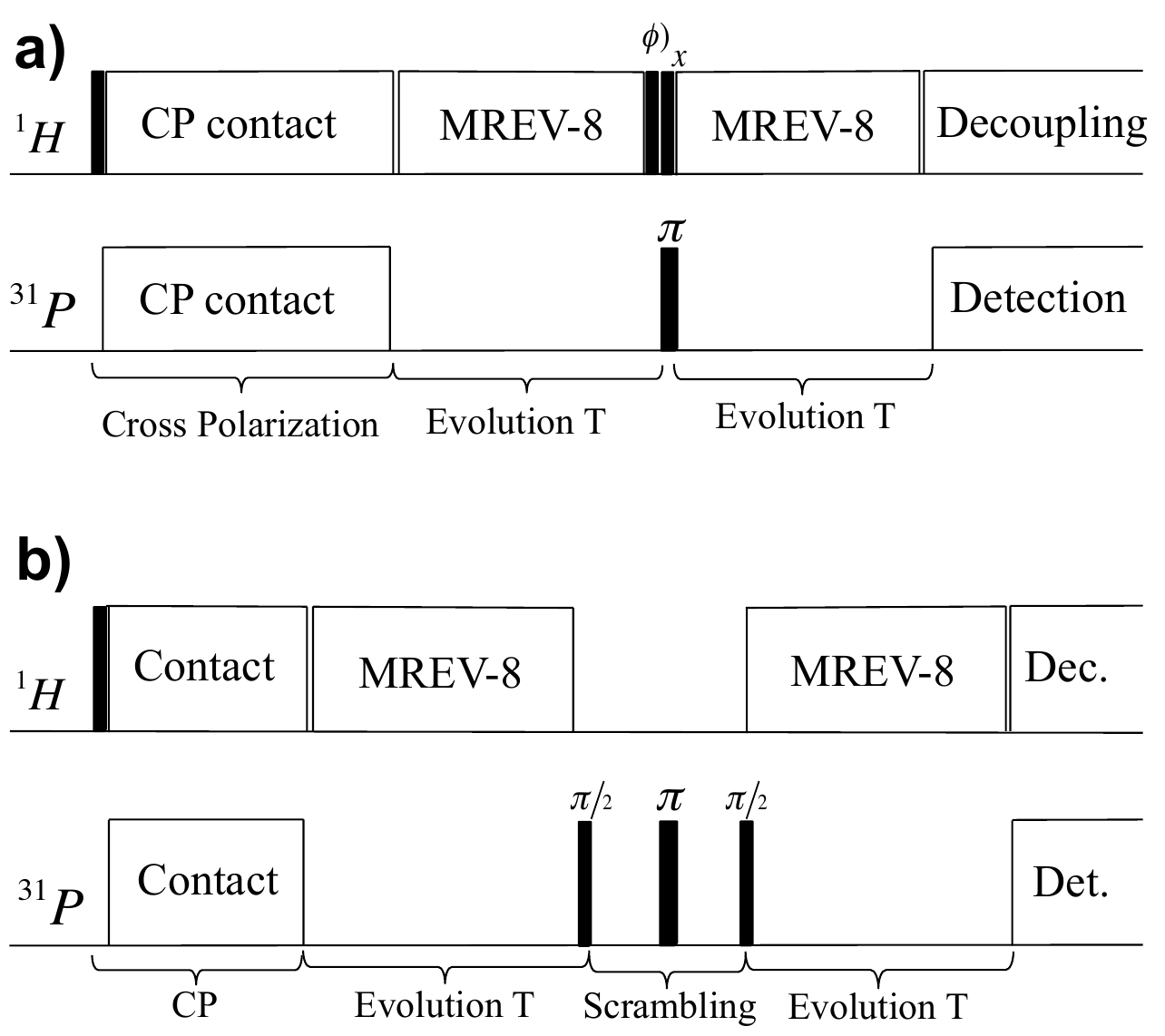} 
	\end{center}	  	 	 	 	
	\caption{The NMR pulse sequence for the multi-spin Correlation Detection (MCD) experiment is shown in panel (a). Panel (b) shows the pulse sequence for the OTOC measurement with quantum information scrambling implementation in the environment.  }
	\label{Fig:NMRexp}
\end{figure}

The MCD experiment captures snapshots of the multi-spin SECs at  specific evolution times. This experiment is designed to initiate the growth of SECs from the central spin, and also to use the central spin itself as a probe for the detection of SECs.  Figure~\ref{Fig:NMRexp}~(a) shows the two channel NMR pulse program used for simultaneous control of the central spin (\textsuperscript{31}$\mathrm{P}$) and the environment spins (\textsuperscript{1}$\mathrm{H}$) in the MCD experiment. The Cross Polarization (CP) step is employed to remove any initial environment correlations, and to increase the sensitivity of the experiment by enhancing the initial polarization of the central spin, in addition to reducing the necessary repetition delay time. 

Evolution under the heteronuclear dipolar interaction for time $ T $ results in the growth of the SECs, while the homonuclear dipolar interaction in the environment is averaged out with the MREV-8 pulse sequence. Under the MREV8 cycle, the  $ \sigmaz $ operator for the environment spins is transformed to a vector pointing at the  $ (1,0,1) $ direction with the scaling factor of~\cite{Rhim73_mrev8}:
\begin{equation}
\alpha= \frac{\sqrt{2}(1+2\frac{3 t_p}{\tau_c} (\frac{4}{\pi}-1))}{3} ,
\end{equation}
where $ t_p $ is the pulse length and $ \tau_c $ is the length of the MREV-8 sequence. Consequently, the zeroth order of the average Hamiltonian for the heteronuclear dipolar interaction in~Eq.~(\ref{hetHam}), in the toggling frame of the MREV-8 pulse sequence, is:
\begin{equation}
\widetilde{\H}_{SE}= 0.36 \sum_{j} \omega_j (  \sigmaz^{cs} \totimes \sigmax^j + \sigmaz^{cs} \totimes \sigmaz^j) .
\end{equation}
Note that because of the symmetry in this Hamiltonian, the $ \sigma_{\text{\tiny X}} $ and $ \sigma_{\text{\tiny Z}} $ operators in the environment are produced with the same weight. Hence, all the equations below Eq.~(\ref{powexp}) are written for  $ \H_{SE} $ instead of $   \widetilde{\H}_{SE} $. This is allowed because this experiment uses $ x $ as the quantization axis, and therefore it is insensitive to $ \sigma_{\text{\tiny X}} $ operators that appear in  $  \widetilde{\H}_{SE} $ and not in $ \H_{SE} $. However, the reader should keep in mind that the environment $ \sigma_{\text{\tiny X}}  $ operators have an equal rate of production as the $ \sigma_{\text{\tiny Z}} $ operators. The only observable difference between these two Hamiltonian in the MCD experiment is that for the $   \widetilde{\H}_{SE} $, the rate of production of the  $ \sigma_{\text{\tiny Z}} $ operators in the environment is scaled down with the scaling factor $ \alpha $. 

As shown in Fig.~\ref{Fig:NMRexp}~(a), after $ T $, we apply  a collective rotation $ \phi_x $ on the environment spins to encode the correlation order as a phase factor $ \e^{i n \phi} $, which is observed at the end of the experiment. Next, the sign of the heteronuclear dipolar interaction is virtually changed by sandwiching the evolution period with $ \pi $ rotations on the central spin, in order to create an echo signal at time $ 2T $. Finally, a decoupling sequence is applied to remove interactions with the environment during the detection and to achieve the maximum signal-to-noise ratio.

Figure~\ref{Fig:NMRexp}~(b) sketches the pulse sequence used for measuring the OTOC decay. In this experiment, the collective rotation of the environment spins is removed and a scrambling window is introduced. During this window, the environment spins evolve under the homonuclear dipolar interaction, while the central spin is decoupled from them. The echo signal at the end of this experiment provides the ratio of the multi-spin correlated terms that were not affected by the homonuclear dipolar interaction during the scrambling window.

\subsection{\label{controlsequences}Control sequence for the environment}

The MREV-8 sequence~\cite{Rhim73_mrev8,HaeberlenBook} is used to freeze the homonuclear dipolar interaction in the environment.  
Under this pulse sequence, the zeroth and the first order terms of the homonuclear dipolar interaction get eliminated and the second order corrections will be the leading factor for line broadening in the environment. The advantage of using the MREV-8 sequence is that pulse imperfections, such as pulse width errors, pulse frequency errors,  and phase transients can be minimized using Average Hamiltonian Theory. The effects of these radio frequency (RF) pulse errors are particularly important in our experiment, since our control sequences are applied repeatedly during the evolution time, and any deviation from the ideal pulse Hamiltonian could accumulate very quickly and potentially destroy the efficiency of the experiment. To take care of remaining errors, we have used a set of experiments known as ``Tune-up Cycles'' for multi-pulse NMR experiments~\cite{Rhim74_tuneupcycles,DybowskiBook}, which enable us to minimize the RF pulse error terms in an iterative process.

\begin{table}[!htb]
	\caption {Characterization of relaxation times with and without the application of the MREV-8 control sequences on the environment spins.  }\label{tab:control}
	\begin{center}
		\begin{tabular}{||c | c||} 
			\hline
			Parameter & Value  \\ [0.5ex] 
			\hline\hline
			Proton T\tsub{1} & 1.5 $\pm$0.1 s \\ 
			Proton T\tsub{2} &  9.8 $\pm 0.2$  $\mu$s \\
			Proton T\tsub{2} with MREV-8     &  8.0 $\pm$0.5  ms   \\
			\hline
			\textsuperscript{31}P T\tsub{1} & 61 $\pm$4 s \\
			\textsuperscript{31}P T\tsub{2}  Hahn echo  & 1.10 $\pm$0.02 ms \\
			\textsuperscript{31}P T\tsub{2} with  MREV-8 on environment & 11.6 $\pm$0.4 ms \\
			\hline
		\end{tabular}
	\end{center}
\end{table}

Table~\ref{tab:control} gives a summary of sample characteristics under our control sequence. The first line for proton and the first line for (\textsuperscript{31}$\mathrm{P}$) indicate the $ T_1 $ relaxation time. Since the initial polarization is transferred from protons to the Phosphorous nuclei, the experiment can be repeated with respect to proton $ T_1 $.

The second line of Table~\ref{tab:control} gives the decoherence time $ T_2 = 9.8\pm 0.2$ $\mu$s associated with the strong homonuclear dipolar couplings in the environment. The application of the MREV-8 sequence removes the majority of the environment interactions and slows down the exchange of information between protons. Therefore, the decay of the NMR signal occurs much slower with decoherence time $ T_2= 8\pm0.5 $ ms (third line of the Table). The dynamics of the central spin also gets affected and the decoherence time for the echo signal of the central spin goes from $ 1.1\pm0.02 $ ms to $ 11.6 \pm 0.4$ ms. 

With the application of the MREV-8 control on the environment spins, the leading terms causing decoherence are the second order terms remaining from the homonuclear dipolar interaction, their cross terms with pulse errors, and also the effects of the external environment. The decoherence time of the central spin reflects the strength of these terms. Notice that even hough decoherence still exists in the sample, the decoherence time of the central spin, $T_2= 11.6 \pm 0.4$ ms, is at least one order of magnitude larger than the evolution time $ 2T=1064 $ $\mu s $. This means that more than $ 91\%  $ of the spin signal survives. This signal drop is taken into account by normalizing the signal amplitude with respect to a reference signal with zero degree encoding pulse. This requires keeping the total length of the experiments precisely constant for each set of experiments. To do this, a composite pulse with constant length is used for the encoding step.

\section{\label{x-basis}Correlations in the $ x $-basis}

\subsection{\label{numberofcorrspins}Growth of the correlated multi-spin terms}

For a closed environment with $ N $ spins, the unitary evolution of the system-environment under the heteronuclear dipolar Hamiltonian, Eq.(\ref{hetHam}), can be written as 
\begin{widetext}
\begin{eqnarray}\label{rhotstaticpic}
\rho(t)&=&U(t).\rho(0).U^{\dagger}(t)\nonumber\\
&=&\frac{1}{2^{N+1}}\{  \sigma^{\text{\tiny CS}}_{\text{\tiny X}}\totimes\mathds{1} ^{ \totimes N} \prod_{i=1}^{N}\cos(\omega_i t )\nonumber \\ 
&+&\sum_{j=1}^N \sigma^{\text{\tiny CS}}_{\text{\tiny Y}}\totimes\sigma^j_{\text{\tiny Z}}\totimes\mathds{1} ^{ \totimes N-1} \sin(\omega_j t)\prod_{i\neq j}^{N}\cos(\omega_i t )\nonumber\\  
&-&\sum_{j,k}^N \sigma^{\text{\tiny CS}}_{\text{\tiny X}}\totimes\sigma^j_{\text{\tiny Z}}\totimes \sigma^k_{\text{\tiny Z}}\totimes\mathds{1} ^{ \totimes N-2} \sin(\omega_j t) \sin(\omega_k t) \prod_{i\neq j,k}^{N}\cos(\omega_i t )\nonumber\\  
&-&\sum_{j,k,l}^N \sigma^{\text{\tiny CS}}_{\text{\tiny Y}}\totimes\sigma^j_{\text{\tiny Z}}\totimes\sigma^k_{\text{\tiny Z}}\totimes\sigma^l_{\text{\tiny Z}}\totimes\mathds{1} ^{ \totimes N-3} \sin(\omega_j t) \sin(\omega_k t)\sin(\omega_l t) \prod_{i\neq j,k,l}^{N}\cos(\omega_i t )\nonumber\\  
&+&\sum_{j,k,l,m}^N\dots \} .
\end{eqnarray} 
\end{widetext}
The equation above can be put in the form of Eq.~(\ref{eq:rhotsigmaz})  if all of the coupling constants are known for the molecules in the ensemble, one can calculate the weight of each correlation order $ \mathscr{C}_n(T) $ in Eq.(\ref{eq:rhotsigmaz}). The equation above shows that higher orders of SECs become non-negligible only at longer evolution times.

In the $ x $-basis the environment part of the total density matrix leads to off-diagonal elements (coherences) along $ x $ axis,  that can be accessed experimentally. In this basis, the ladder operators are $\Sigma^{j}_{\pm} =\sigmay^j \pm i \sigmaz^j$ and the density matrix is written as
\begin{widetext}
\begin{eqnarray}
 \hspace{-0. cm}\rho{(T)} &=&C_0(T) \sum_{j\neq k}^{N} \sigmax^{\text{cs}} \totimes [\mathds{1}^{N}-(\Sigma^j_{+}\totimes\Sigma^k_{-}\totimes \mathds{1}^{N-2}) +\cdots ] \\
 \hspace{-0. cm} &+&C_1(T) \sum_{j}^{N}\sigmay^{\text{cs}}  \totimes [(\Sigma^j_{+}\totimes \mathds{1}^{N-1})-(\Sigma^j_{-} \totimes \mathds{1}^{N-1})
+\cdots] \nonumber\\
 \hspace{-0. cm} &+& C_2(T) \sum_{j\neq k}^{N}\sigmax^{\text{cs}}  \totimes [(\Sigma^j_{+}\totimes \Sigma^k_{+}\totimes \mathds{1}^{N-2}) +(\Sigma^j_{-}\totimes \Sigma^k_{-} \totimes \mathds{1}^{N-2})+\cdots]\nonumber\\
 \hspace{-0. cm} &+&\cdots   \nonumber \\
 \vspace{0.5 cm} \nonumber\\
 &=&\sum_{n} C_n(T) \rho_{n}^{\text x} \nonumber
\label{eq:rhotalpha3x}
\end{eqnarray}
\end{widetext}
$ \rho_{n}^{\text x} $'s are vectors from the Liouville space describing the measurement basis of the MCD experiment, and they include all permutations of $ \Sigma_{\pm} $ operators with correlation order $ n $.

\subsection{\label{xquantizationcorr}Correlation orders vs number of correlated spins}

To understand the spin physics of the MCD experiment, an example of a central spin model with two spins in the environment, $ N=2 $, is explored in this section. After the cross polarization step, the density matrix for the central spin and the environment is
\begin{equation}
\rho(0)=\frac{1}{2^3}   \sigma^{\text{\tiny CS}}_{\text{\tiny X}} \totimes \mathds{1} \totimes \mathds{1} . 
\end{equation}
Since pre-existing correlations between the central spin and the environment spins disappear during the spin locking pulse, the initial system-environment state is uncorrelated. In order to maintain our SEC terms as simple as possible,  we assume that the homonuclear dipolar interaction in the environment is completely turned off during the evolution step and the central spin evolves under the heteronuclear dipolar interaction:
\begin{equation}\label{eq:Hhet12}
\H^{1,2}_{SE}= \frac{\omega_1}{2}  \{\sigma^{\text{\tiny CS}}_{\text{\tiny Z}} \totimes \sigma^1_{\text{\tiny Z}}\totimes \mathds{1}\}+\frac{\omega_2}{2} \{\sigma^{\text{\tiny CS}}_{\text{\tiny Z}}\totimes \mathds{1} \totimes \sigma^2_{\text{\tiny Z}}\}.
\end{equation}
After the evolution time $ T $, the density matrix evolves to
\begin{eqnarray}	
\rho(T)= \frac{1}{8} &\Big\{& \cos(\omega_1 T)\cos(\omega_2 T) \quad \sigma^{\text{\tiny CS}}_{\text{\tiny X}} \totimes \mathds{1} \totimes \mathds{1} \label{3spinrhoT}\\
&+& \sin(\omega_1 T) \cos(\omega_2 T) \quad\sigma^{\text{\tiny CS}}_{\text{\tiny Y}}\totimes\sigma_{\text{\tiny Z}}\totimes\mathds{1} \nonumber\\
&+&\cos(\omega_1 T) \sin(\omega_2 T)\quad \sigma^{\text{\tiny CS}}_{\text{\tiny Y}}\totimes\mathds{1}\totimes\sigma_{\text{\tiny Z}}  \nonumber\\
&-& \sin(\omega_1 T) \sin(\omega_2 T)\quad \sigma^{\text{\tiny CS}}_{\text{\tiny X}} \totimes\sigma_{\text{\tiny Z}}\totimes\sigma_{\text{\tiny Z}}  \Big\} .\nonumber
\end{eqnarray}    
This is equivalent to the description of the density matrix using the number of coupled spins, Eq.(\ref{eq:rhotsigmaz}). The coefficients of the various spin terms above correspond to the $ \mathscr C_n(T) $'s in Eq.(\ref{eq:rhotsigmaz}). Notice that in the NMR experiment, spins are not distinguishable and only the sum of all single spin correlation terms in Eq.(\ref{3spinrhoT}) are observed.  Collective rotation of the environment spins by $ \phi$ about the $ x $ axis, $ R_{\text{\tiny X}}(\phi)=\exp(i \frac{\phi}{2}\sum_i \mathds{1}^{\text{\tiny CS}}\totimes \sigma^i_{\text{\tiny X}} ) $,  transforms the density matrix to:
\begin{widetext}
\begin{eqnarray}
\rho_{\phi}(T)&=& \frac{1}{8} \Big\{\sigma^{\text{\tiny CS}}_{\text{\tiny X}} \totimes \mathds{1} \totimes \mathds{1} \cos(\omega_1 T)\cos(\omega_2 T) \\
&+&\cos(\phi)\{\sigma^{\text{\tiny CS}}_{\text{\tiny Y}}\totimes\sigma_{\text{\tiny Z}}\totimes\mathds{1} \sin(\omega_1 T) \cos(\omega_2 T)+ \sigma^{\text{\tiny CS}}_{\text{\tiny Y}}\totimes\mathds{1}\totimes\sigma_{\text{\tiny Z}} \cos(\omega_1 T) \sin(\omega_2 T) \}\nonumber\\
&+&\sin(\phi)\{\sigma^{\text{\tiny CS}}_{\text{\tiny Y}}\totimes\sigma_{\text{\tiny Y}}\totimes\mathds{1} \sin(\omega_1 T) \cos(\omega_2 T) + \sigma^{\text{\tiny CS}}_{\text{\tiny Y}}\totimes\mathds{1}\totimes\sigma_{\text{\tiny Y}} \cos(\omega_1 T) \sin(\omega_2 T) \} \nonumber\\
&-&\sin(\omega_1 T) \sin(\omega_2 T) \{\cos(\phi)^2 \sigma^{\text{\tiny CS}}_{\text{\tiny X}} \totimes\sigma_{\text{\tiny Z}}\totimes\sigma_{\text{\tiny Z}}  +\sin(\phi)^2 \sigma^{\text{\tiny CS}}_{\text{\tiny X}} \totimes\sigma_{\text{\tiny Y}}\totimes\sigma_{\text{\tiny Y}}\} \nonumber\\
&-& \sin(\omega_1 T) \sin(\omega_2 T) \cos(\phi)\sin(\phi)\{ \sigma^{\text{\tiny CS}}_{\text{\tiny X}} \totimes\sigma_{\text{\tiny Z}}\totimes\sigma_{\text{\tiny Y}}  + \sigma^{\text{\tiny CS}}_{\text{\tiny X}} \totimes\sigma_{\text{\tiny Y}}\totimes\sigma_{\text{\tiny Z}}\}  \Big\} . \nonumber
\end{eqnarray} 
\end{widetext}
Consequently, the density matrix terms  gain a $ \cos(\phi)^n $ factor where $ n $ corresponds to the number of  $ \sigmaz $ operators in the multi-spin correlated terms. The next step is  another evolution interval $ T $ with the inverse of Eq.(\ref{eq:Hhet12}). The resulting density matrix $ \rho_{\phi}(2T) $ is given by a long equation shown in  Ref~\cite{Mohamad_PhDthesis}. But from this equation, the only observable terms are the following ones
\begin{gather*}
\begin{cases}
\qquad \cos(\omega_1 T)^2 \cos(\omega_2 T)^2 \quad\sigmax^{\text{\tiny CS}} \totimes \mathds{1} \totimes \mathds{1} \\
\cos(\phi) \sin(\omega_1 T)^2 \cos(\omega_2 T)^2\quad \sigmax^{\text{\tiny CS}}\totimes\mathds{1}\totimes\mathds{1}\\
\cos(\phi) \cos(\omega_1 T)^2 \sin(\omega_2 T)^2 \quad \sigmax^{\text{\tiny CS}}\totimes\mathds{1}\totimes\mathds{1}\\
\cos(\phi)^2 \sin(\omega_1 T)^2 \sin(\omega_2 T)^2 \quad \sigmax^{\text{\tiny CS}}\totimes\mathds{1}\totimes\mathds{1}
\end{cases}
\end{gather*}
The signal amplitude is evaluated with the inner product of the reduced state of the central spin and the measurement operator, $ \sigma^{\text{\tiny CS}}_{\text{\tiny X}} $, at  $ 2T $:
\begin{eqnarray}
S_{\phi}(2T)&=& \Tr[\Tr_{\text{E}}[\rho(2T)] .\sigma^{\text{\tiny CS}}_{\text{\tiny X}}]\\
&=&\cos(\omega_1 T)^2\cos(\omega_2 T)^2\label{3spinrho2T}\nonumber\\
&+&\cos(\phi)\{\cos(\omega_1 T)^2\sin(\omega_2 T)^2\nonumber\\
&&\qquad +\sin(\omega_1 T)^2\cos(\omega_2 T)^2\} \nonumber\\
&+&\cos(\phi)^2 \sin(\omega_1 T)^2\sin(\omega_2 T)^2 .\nonumber
\end{eqnarray}
The data set containing amplitudes of $ S_{\phi}(2T) $ for various encoding angles $ \phi $ is Fourier transformed to evaluate the weight of each correlation order $ |C_n(T)|^2 $: 
\begin{widetext}
\begin{eqnarray}
\mathbb{F}[S_{\phi}(2T)]&=& \cos(\omega_1 T)^2\cos(\omega_2 T)^2 \delta(n)\\
&+&\{\cos(\omega_1 T)^2\sin(\omega_2 T)^2+\sin(\omega_1 T)^2\cos(\omega_2 T)^2\} [\frac{1}{2}\delta(n-1)+\frac{1}{2}\delta(n+1)]\nonumber\\
&+& \sin(\omega_1 T)^2\sin(\omega_2 T)^2 [\frac{1}{4} \delta(n-2)+\frac{1}{2}\delta(n)+\frac{1}{4}\delta(n+2)] .\nonumber
\end{eqnarray}
\end{widetext}
We can compare these coefficients with the density matrix weights $ \mathscr C_n (T)$ expressed in Eq.(\ref{3spinrhoT}). First, notice that the amplitude of the order $ n $ of the Fourier transformed signal is given by the squared coefficients $ |\mathscr C_n(T)|^2 $ of   $ \rho(T) $. The first line in the equation above  is the signal resulted from the uncorrelated spin term in the first line of Eq.(\ref{3spinrhoT}), the $\mathscr C_0(T) $ term. In the SEC spectrum, this is the amplitude for $ n=0 $.  The second line is produced by spin terms with one spin correlated to the central spin, that is the $\mathscr C_1(T) $ terms. In the SEC spectrum they show up at $ n=\pm 1 $. The third line is produced by the term which has two environment spins correlated to the central spin, that is the  $\mathscr C_2(T) $ term. In the SEC spectrum it shows up at $ n=0,\pm2 $. Thus, the $ \mathscr C_2(T) $ term contributes to the production of both $ C_0(T) $ and $ C_2(T) $ terms. This explains why in Fig.~\ref{Fig:MCD}~(c), $ |C_0(T)|^2 $ and $ |C_2(T)|^2 $ have similar values after decay of  the $ \mathscr C_0(T) $ term and before the $ \mathscr C_4(T) $ term becomes significant.

It is easy to show that all even (odd) powers of $ \cos(\phi)^n $ in the $ \mathscr C_n(T) $ terms produce Fourier components at  even(odd) orders of $ C_n(T)$ terms,  where the amplitude of each peak is evaluated with the coefficients of the binomial distribution~\cite{Mohamad_PhDthesis}. Therefore,  the amplitude of the largest observed correlation order for each molecule scales down with a factor of $ \frac{1}{2^n} $. Consequently, the second moment of the correlation order spectrum is more suitable as a measure for the extent of SECs, than the largest observed order.

\section{\label{coingame}Classical coin game}

\begin{figure}[!bth]
	\begin{center}
		\includegraphics[scale=0.5]{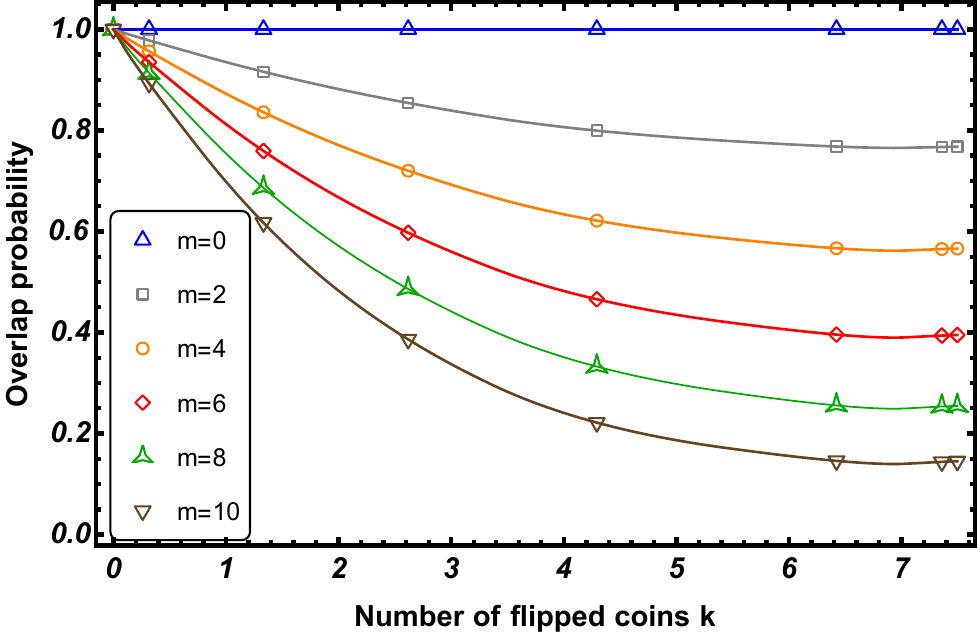} 
	\end{center}	  	 	 	 	
	\caption{The classical coin game is similar to the scrambling of quantum information in the spin environment. The overlap between the initial and final state of the coin array decays exponentially with the number of performed coin swaps.}
	\label{Fig:coinsim}
\end{figure}

We have designed a classical game to make a parallel with the swap dynamics of the environment spins that take place in our second experiment. This game simulates the loss of echo signal  resulting only from spin swaps between  the connected and  non-connected spin groups. It does not address the decay resulting from spin swaps in the connected spin group. Consider an array of $ N $ coins initially set to heads. We randomly flip $ k $ of these coins  to represent spins in the connected group at time $ T $, with the constraint that each coin may be flipped only once. The remaining $ N-k $ coins represent the non-connected group. Subsequently, if the coins are not swapped, flipping the same random $ k $ coins for a second time results in the complete return to the initial state. This is equivalent to a perfect echo of the spin signal at time $ 2T $. However, when we add random swap actions between the two flipping stages, the final state of the coin array may be different from its initial state. The distance between the initial and the final state of the coin array depends on the number of swaps performed between  flipped coins and  un-flipped coins. The probability of having this sort of ``successful swap'' (ssw) for each spin pair, that is swaps that increase the distance between the two states of the coin array, is given by
\begin{equation}
P_{ssw}= 2\frac{kN-k^2}{N^2-N} \cdot \label{Psswap}
\end{equation} 
We  ignore the cases where the same two coins swap more than once, and we assume that after each swap, the probability of having a successful swap remains unchanged. Then the probability of success for $ m $ random coin swaps is $ m $ times the probability of success for one coin swap. Consequently, the overlap amplitude $ A_{OL} $ between the initial state and the final state of the coin chain after two rounds of flips with a round of coin swap in the middle is 
\begin{eqnarray}
A_{OL}(m,k,N)&:=& \left(1- \frac{2m}{N}P_{ssw}\right)^2 \nonumber\\
&=& \left(1-\frac{4m}{N}\frac{kN-k^2}{N^2-N}\right)^2 . \label{CEA}
\end{eqnarray}         
$ A_{OL}(m,k,N) $ is plotted in Fig.~\ref{Fig:coinsim} for an array of $ N=15 $ coins, which is the number of environment spins, while $ k $ is set according to the  Hamming weight spread for various evolution time steps $ T $ in the MCD experiment, and $ m $ is varied from 0 to 10.

\begin{figure}[!htb]
	\begin{center}
		\includegraphics[scale=0.5]{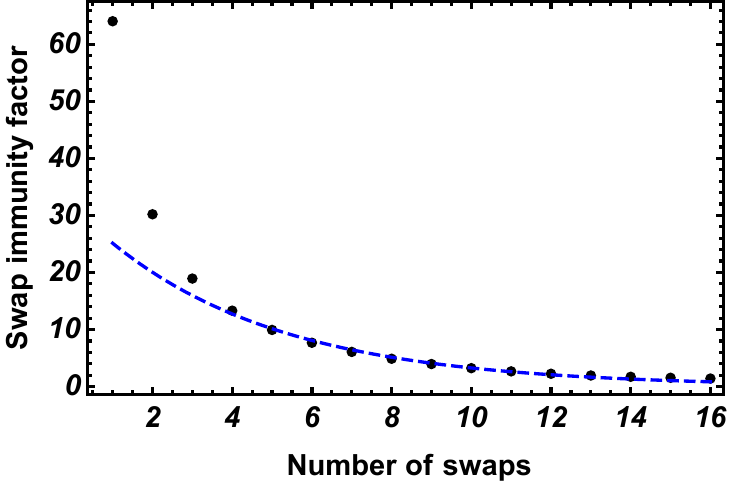} 
	\end{center}	  	 	 	 	
	\caption{The swap immunity factor for $ m>4 $ shows an exponentially decaying behavior, similar to the experimental results for the Fig.\ref{Fig:OTOC}~(c). The dashed line indicates an exponential fit for this part of the data. For $ m<4 $ the swap immunity factor is larger than the size of the coin array, which means that coin swap cannot effectively disturb the coin array recovery. This final remark is also similar to the discussion about Fig.\ref{Fig:OTOC}~(c), when $ \tau $ is small. }
	\label{Fig:coineffsize}
\end{figure}

We have fitted the data in Fig.\ref{Fig:coinsim} with a series of decaying exponential functions to characterize the capability of various numbers of coin swaps to disrupt the overlap between the initial and final coin state~\cite{Mohamad_PhDthesis}. The overlap probability becomes $ \frac{1}{e} $, when $ k  $ is the inverse of the exponential decay rate. We call this inverse rate the
``swap immunity factor'' and plot it as a function of $ m $ in Fig.~\ref{Fig:coineffsize}. The swap immunity factor indicates the effectiveness of the number of coin swaps in obstructing the coin array recovery. Similar to the non-local OTOC experimental results in Fig.~\ref{Fig:OTOC}~(c), the  swap immunity factor decays exponentially with $ m $, for $ m $ above a threshold. 

\section{\label{qchaos}Chaotic environment}
Quantum chaos refers to properties of the spectrum that indicate whether the classical counterpart of the quantum system is chaotic. One of the main signatures of chaos is the strong repulsion of the eigenvalues~\cite{Guhr1998}. The energy levels of quantum systems that are classically chaotic are correlated and prohibited from crossing. This is detected, for example, with the distribution $P(s)$ of the unfolded spacings $s$ between neighboring levels. In the case of real and symmetric Hamiltonian matrices, as in our case, the level spacing distribution follows closely the Wigner surmise,
\begin{equation}
P(s)= \frac{\pi s}{2} \exp \left(- \frac{\pi s^2}{4} \right).
\end{equation}
We verified that $ P(s) $ for the Hamiltonian in Eq.~(\ref{Eq:bath}) is well described with this equation. The spread of information in chaotic systems far from equilibrium happens very fast~\cite{Borgonovi2016}. This is the scenario of our composite system, where information is initially confined to a  single spin and the environment is chaotic.


\begin{thebibliography}{55}%
	\makeatletter
	\providecommand \@ifxundefined [1]{%
		\@ifx{#1\undefined}
	}%
	\providecommand \@ifnum [1]{%
		\ifnum #1\expandafter \@firstoftwo
		\else \expandafter \@secondoftwo
		\fi
	}%
	\providecommand \@ifx [1]{%
		\ifx #1\expandafter \@firstoftwo
		\else \expandafter \@secondoftwo
		\fi
	}%
	\providecommand \natexlab [1]{#1}%
	\providecommand \enquote  [1]{``#1''}%
	\providecommand \bibnamefont  [1]{#1}%
	\providecommand \bibfnamefont [1]{#1}%
	\providecommand \citenamefont [1]{#1}%
	\providecommand \href@noop [0]{\@secondoftwo}%
	\providecommand \href [0]{\begingroup \@sanitize@url \@href}%
	\providecommand \@href[1]{\@@startlink{#1}\@@href}%
	\providecommand \@@href[1]{\endgroup#1\@@endlink}%
	\providecommand \@sanitize@url [0]{\catcode `\\12\catcode `\$12\catcode
		`\&12\catcode `\#12\catcode `\^12\catcode `\_12\catcode `\%12\relax}%
	\providecommand \@@startlink[1]{}%
	\providecommand \@@endlink[0]{}%
	\providecommand \url  [0]{\begingroup\@sanitize@url \@url }%
	\providecommand \@url [1]{\endgroup\@href {#1}{\urlprefix }}%
	\providecommand \urlprefix  [0]{URL }%
	\providecommand \Eprint [0]{\href }%
	\providecommand \doibase [0]{https://doi.org/}%
	\providecommand \selectlanguage [0]{\@gobble}%
	\providecommand \bibinfo  [0]{\@secondoftwo}%
	\providecommand \bibfield  [0]{\@secondoftwo}%
	\providecommand \translation [1]{[#1]}%
	\providecommand \BibitemOpen [0]{}%
	\providecommand \bibitemStop [0]{}%
	\providecommand \bibitemNoStop [0]{.\EOS\space}%
	\providecommand \EOS [0]{\spacefactor3000\relax}%
	\providecommand \BibitemShut  [1]{\csname bibitem#1\endcsname}%
	\let\auto@bib@innerbib\@empty
	\bibitem [{\citenamefont {Bennett}\ and\ \citenamefont
		{DiVincenzo}(2000)}]{Bennett2000}%
	\BibitemOpen
	\bibfield  {author} {\bibinfo {author} {\bibfnamefont {C.~H.}\ \bibnamefont
			{Bennett}}\ and\ \bibinfo {author} {\bibfnamefont {D.~P.}\ \bibnamefont
			{DiVincenzo}},\ }\bibfield  {title} {\bibinfo {title} {Quantum information
			and computation},\ }\href {http://dx.doi.org/10.1038/35005001} {\bibfield
		{journal} {\bibinfo  {journal} {Nature}\ }\textbf {\bibinfo {volume} {404}},\
		\bibinfo {pages} {247} (\bibinfo {year} {2000})}\BibitemShut {NoStop}%
	\bibitem [{\citenamefont {Myatt}\ \emph {et~al.}(2000)\citenamefont {Myatt},
		\citenamefont {King}, \citenamefont {Turchette}, \citenamefont {Sackett},
		\citenamefont {Kielpinski}, \citenamefont {Itano}, \citenamefont {Monroe},\
		and\ \citenamefont {Wineland}}]{Myatt2000}%
	\BibitemOpen
	\bibfield  {author} {\bibinfo {author} {\bibfnamefont {C.~J.}\ \bibnamefont
			{Myatt}}, \bibinfo {author} {\bibfnamefont {B.~E.}\ \bibnamefont {King}},
		\bibinfo {author} {\bibfnamefont {Q.~A.}\ \bibnamefont {Turchette}}, \bibinfo
		{author} {\bibfnamefont {C.~A.}\ \bibnamefont {Sackett}}, \bibinfo {author}
		{\bibfnamefont {D.}~\bibnamefont {Kielpinski}}, \bibinfo {author}
		{\bibfnamefont {W.~M.}\ \bibnamefont {Itano}}, \bibinfo {author}
		{\bibfnamefont {C.}~\bibnamefont {Monroe}},\ and\ \bibinfo {author}
		{\bibfnamefont {D.~J.}\ \bibnamefont {Wineland}},\ }\bibfield  {title}
	{\bibinfo {title} {Decoherence of quantum superpositions through coupling to
			engineered reservoirs},\ }\href {http://dx.doi.org/10.1038/35002001}
	{\bibfield  {journal} {\bibinfo  {journal} {Nature}\ }\textbf {\bibinfo
			{volume} {403}},\ \bibinfo {pages} {269} (\bibinfo {year}
		{2000})}\BibitemShut {NoStop}%
	\bibitem [{\citenamefont {Khaetskii}\ \emph {et~al.}(2002)\citenamefont
		{Khaetskii}, \citenamefont {Loss},\ and\ \citenamefont
		{Glazman}}]{Khaetskii02}%
	\BibitemOpen
	\bibfield  {author} {\bibinfo {author} {\bibfnamefont {A.~V.}\ \bibnamefont
			{Khaetskii}}, \bibinfo {author} {\bibfnamefont {D.}~\bibnamefont {Loss}},\
		and\ \bibinfo {author} {\bibfnamefont {L.}~\bibnamefont {Glazman}},\
	}\bibfield  {title} {\bibinfo {title} {Electron spin decoherence in quantum
			dots due to interaction with nuclei},\ }\href
	{https://doi.org/10.1103/PhysRevLett.88.186802} {\bibfield  {journal}
		{\bibinfo  {journal} {Phys. Rev. Lett.}\ }\textbf {\bibinfo {volume} {88}},\
		\bibinfo {pages} {186802} (\bibinfo {year} {2002})}\BibitemShut {NoStop}%
	\bibitem [{\citenamefont {Coish}\ and\ \citenamefont {Loss}(2004)}]{Coish2004}%
	\BibitemOpen
	\bibfield  {author} {\bibinfo {author} {\bibfnamefont {W.~A.}\ \bibnamefont
			{Coish}}\ and\ \bibinfo {author} {\bibfnamefont {D.}~\bibnamefont {Loss}},\
	}\bibfield  {title} {\bibinfo {title} {Hyperfine interaction in a quantum
			dot: Non-{Markovian} electron spin dynamics},\ }\href
	{https://doi.org/10.1103/PhysRevB.70.195340} {\bibfield  {journal} {\bibinfo
			{journal} {Phys. Rev. B}\ }\textbf {\bibinfo {volume} {70}},\ \bibinfo
		{pages} {195340} (\bibinfo {year} {2004})}\BibitemShut {NoStop}%
	\bibitem [{\citenamefont {Cucchietti}\ \emph {et~al.}(2005)\citenamefont
		{Cucchietti}, \citenamefont {Paz},\ and\ \citenamefont
		{Zurek}}]{Cucchietti2005}%
	\BibitemOpen
	\bibfield  {author} {\bibinfo {author} {\bibfnamefont {F.~M.}\ \bibnamefont
			{Cucchietti}}, \bibinfo {author} {\bibfnamefont {J.~P.}\ \bibnamefont
			{Paz}},\ and\ \bibinfo {author} {\bibfnamefont {W.~H.}\ \bibnamefont
			{Zurek}},\ }\bibfield  {title} {\bibinfo {title} {Decoherence from spin
			environments},\ }\href {https://doi.org/10.1103/PhysRevA.72.052113}
	{\bibfield  {journal} {\bibinfo  {journal} {Phys. Rev. A}\ }\textbf {\bibinfo
			{volume} {72}},\ \bibinfo {pages} {052113} (\bibinfo {year}
		{2005})}\BibitemShut {NoStop}%
	\bibitem [{\citenamefont {Yang}\ and\ \citenamefont {Liu}(2008)}]{Yang2008}%
	\BibitemOpen
	\bibfield  {author} {\bibinfo {author} {\bibfnamefont {W.}~\bibnamefont
			{Yang}}\ and\ \bibinfo {author} {\bibfnamefont {R.-B.}\ \bibnamefont {Liu}},\
	}\bibfield  {title} {\bibinfo {title} {Quantum many-body theory of qubit
			decoherence in a finite-size spin bath},\ }\href
	{http://link.aps.org/doi/10.1103/PhysRevB.78.085315} {\bibfield  {journal}
		{\bibinfo  {journal} {Physical Review B}\ }\textbf {\bibinfo {volume} {78}},\
		\bibinfo {pages} {085315} (\bibinfo {year} {2008})}\BibitemShut {NoStop}%
	\bibitem [{\citenamefont {Pernice}\ \emph {et~al.}(2012)\citenamefont
		{Pernice}, \citenamefont {Helm},\ and\ \citenamefont {Strunz}}]{Pernice2012}%
	\BibitemOpen
	\bibfield  {author} {\bibinfo {author} {\bibfnamefont {A.}~\bibnamefont
			{Pernice}}, \bibinfo {author} {\bibfnamefont {J.}~\bibnamefont {Helm}},\ and\
		\bibinfo {author} {\bibfnamefont {W.~T.}\ \bibnamefont {Strunz}},\ }\bibfield
	{title} {\bibinfo {title} {System--environment correlations and
			non-{Markovian} dynamics},\ }\href
	{http://stacks.iop.org/0953-4075/45/i=15/a=154005} {\bibfield  {journal}
		{\bibinfo  {journal} {Journal of Physics B: Atomic, Molecular and Optical
				Physics}\ }\textbf {\bibinfo {volume} {45}},\ \bibinfo {pages} {154005}
		(\bibinfo {year} {2012})}\BibitemShut {NoStop}%
	\bibitem [{\citenamefont {Ma}\ \emph {et~al.}(2014)\citenamefont {Ma},
		\citenamefont {Wolfowicz}, \citenamefont {Zhao}, \citenamefont {Li},
		\citenamefont {Morton},\ and\ \citenamefont {Liu}}]{Ma2014}%
	\BibitemOpen
	\bibfield  {author} {\bibinfo {author} {\bibfnamefont {W.-L.}\ \bibnamefont
			{Ma}}, \bibinfo {author} {\bibfnamefont {G.}~\bibnamefont {Wolfowicz}},
		\bibinfo {author} {\bibfnamefont {N.}~\bibnamefont {Zhao}}, \bibinfo {author}
		{\bibfnamefont {S.-S.}\ \bibnamefont {Li}}, \bibinfo {author} {\bibfnamefont
			{J.~J.}\ \bibnamefont {Morton}},\ and\ \bibinfo {author} {\bibfnamefont
			{R.-B.}\ \bibnamefont {Liu}},\ }\bibfield  {title} {\bibinfo {title}
		{Uncovering many-body correlations in nanoscale nuclear spin baths by central
			spin decoherence},\ }\href {https://doi.org/10.1038/ncomms5822} {\bibfield
		{journal} {\bibinfo  {journal} {Nature Communications}\ }\textbf {\bibinfo
			{volume} {5}},\ \bibinfo {pages} {4822} (\bibinfo {year} {2014})}\BibitemShut
	{NoStop}%
	\bibitem [{\citenamefont {Breuer}\ \emph {et~al.}(2009)\citenamefont {Breuer},
		\citenamefont {Laine},\ and\ \citenamefont {Piilo}}]{Breuer2009}%
	\BibitemOpen
	\bibfield  {author} {\bibinfo {author} {\bibfnamefont {H.-P.}\ \bibnamefont
			{Breuer}}, \bibinfo {author} {\bibfnamefont {E.-M.}\ \bibnamefont {Laine}},\
		and\ \bibinfo {author} {\bibfnamefont {J.}~\bibnamefont {Piilo}},\ }\bibfield
	{title} {\bibinfo {title} {Measure for the degree of non-{Markovian}
			behavior of quantum processes in open systems},\ }\href
	{https://doi.org/10.1103/PhysRevLett.103.210401} {\bibfield  {journal}
		{\bibinfo  {journal} {Phys. Rev. Lett.}\ }\textbf {\bibinfo {volume} {103}},\
		\bibinfo {pages} {210401} (\bibinfo {year} {2009})}\BibitemShut {NoStop}%
	\bibitem [{\citenamefont {Breuer}\ \emph {et~al.}(2016)\citenamefont {Breuer},
		\citenamefont {Laine}, \citenamefont {Piilo},\ and\ \citenamefont
		{Vacchini}}]{Breuer2016}%
	\BibitemOpen
	\bibfield  {author} {\bibinfo {author} {\bibfnamefont {H.-P.}\ \bibnamefont
			{Breuer}}, \bibinfo {author} {\bibfnamefont {E.-M.}\ \bibnamefont {Laine}},
		\bibinfo {author} {\bibfnamefont {J.}~\bibnamefont {Piilo}},\ and\ \bibinfo
		{author} {\bibfnamefont {B.}~\bibnamefont {Vacchini}},\ }\bibfield  {title}
	{\bibinfo {title} {Colloquium: Non-{Markovian} dynamics in open quantum
			systems},\ }\href {https://doi.org/10.1103/RevModPhys.88.021002} {\bibfield
		{journal} {\bibinfo  {journal} {Rev. Mod. Phys.}\ }\textbf {\bibinfo {volume}
			{88}},\ \bibinfo {pages} {021002} (\bibinfo {year} {2016})}\BibitemShut
	{NoStop}%
	\bibitem [{\citenamefont {Baum}\ \emph {et~al.}(1985)\citenamefont {Baum},
		\citenamefont {Munowitz}, \citenamefont {Garroway},\ and\ \citenamefont
		{Pines}}]{Baum1985}%
	\BibitemOpen
	\bibfield  {author} {\bibinfo {author} {\bibfnamefont {J.}~\bibnamefont
			{Baum}}, \bibinfo {author} {\bibfnamefont {M.}~\bibnamefont {Munowitz}},
		\bibinfo {author} {\bibfnamefont {A.~N.}\ \bibnamefont {Garroway}},\ and\
		\bibinfo {author} {\bibfnamefont {A.}~\bibnamefont {Pines}},\ }\bibfield
	{title} {\bibinfo {title} {Multiple-quantum dynamics in solid state
			{{{NMR}}}},\ }\href {https://doi.org/10.1063/1.449344} {\bibfield  {journal}
		{\bibinfo  {journal} {J. Chem. Phys.}\ }\textbf {\bibinfo {volume} {83}},\
		\bibinfo {pages} {2015} (\bibinfo {year} {1985})}\BibitemShut {NoStop}%
	\bibitem [{\citenamefont {Munowitz}\ \emph {et~al.}(1987)\citenamefont
		{Munowitz}, \citenamefont {Pines},\ and\ \citenamefont
		{Mehring}}]{Munowitz1987}%
	\BibitemOpen
	\bibfield  {author} {\bibinfo {author} {\bibfnamefont {M.}~\bibnamefont
			{Munowitz}}, \bibinfo {author} {\bibfnamefont {A.}~\bibnamefont {Pines}},\
		and\ \bibinfo {author} {\bibfnamefont {M.}~\bibnamefont {Mehring}},\
	}\bibfield  {title} {\bibinfo {title} {Multiple-quantum dynamics in {NMR}: A
			directed walk through {Liouville} space},\ }\href
	{https://doi.org/10.1063/1.452028} {\bibfield  {journal} {\bibinfo  {journal}
			{J. Chem. Phys.}\ }\textbf {\bibinfo {volume} {86}},\ \bibinfo {pages} {3172}
		(\bibinfo {year} {1987})}\BibitemShut {NoStop}%
	\bibitem [{\citenamefont {Lacelle}\ \emph {et~al.}(1993)\citenamefont
		{Lacelle}, \citenamefont {Hwang},\ and\ \citenamefont
		{Gerstein}}]{Lacelle1993}%
	\BibitemOpen
	\bibfield  {author} {\bibinfo {author} {\bibfnamefont {S.}~\bibnamefont
			{Lacelle}}, \bibinfo {author} {\bibfnamefont {S.-J.}\ \bibnamefont {Hwang}},\
		and\ \bibinfo {author} {\bibfnamefont {B.~C.}\ \bibnamefont {Gerstein}},\
	}\bibfield  {title} {\bibinfo {title} {Multiple quantum nuclear magnetic
			resonance of solids: A cautionary note for data analysis and
			interpretation},\ }\bibfield  {booktitle} {\emph {\bibinfo {booktitle} {The
				Journal of Chemical Physics}},\ }\href {https://doi.org/10.1063/1.465616}
	{\bibfield  {journal} {\bibinfo  {journal} {J. Chem. Phys.}\ }\textbf
		{\bibinfo {volume} {99}},\ \bibinfo {pages} {8407} (\bibinfo {year}
		{1993})}\BibitemShut {NoStop}%
	\bibitem [{\citenamefont {Ramanathan}\ \emph {et~al.}(2003)\citenamefont
		{Ramanathan}, \citenamefont {Cho}, \citenamefont {Cappellaro}, \citenamefont
		{Boutis},\ and\ \citenamefont {Cory}}]{Ramanathan2003}%
	\BibitemOpen
	\bibfield  {author} {\bibinfo {author} {\bibfnamefont {C.}~\bibnamefont
			{Ramanathan}}, \bibinfo {author} {\bibfnamefont {H.}~\bibnamefont {Cho}},
		\bibinfo {author} {\bibfnamefont {P.}~\bibnamefont {Cappellaro}}, \bibinfo
		{author} {\bibfnamefont {G.~S.}\ \bibnamefont {Boutis}},\ and\ \bibinfo
		{author} {\bibfnamefont {D.~G.}\ \bibnamefont {Cory}},\ }\bibfield  {title}
	{\bibinfo {title} {Encoding multiple quantum coherences in non-commuting
			bases},\ }\href
	{https://doi.org/http://dx.doi.org/10.1016/S0009-2614(02)02020-1} {\bibfield
		{journal} {\bibinfo  {journal} {Chem. Phys. Lett.}\ }\textbf {\bibinfo
			{volume} {369}},\ \bibinfo {pages} {311} (\bibinfo {year}
		{2003})}\BibitemShut {NoStop}%
	\bibitem [{\citenamefont {Krojanski}\ and\ \citenamefont
		{Suter}(2004)}]{Krojanski2004}%
	\BibitemOpen
	\bibfield  {author} {\bibinfo {author} {\bibfnamefont {H.~G.}\ \bibnamefont
			{Krojanski}}\ and\ \bibinfo {author} {\bibfnamefont {D.}~\bibnamefont
			{Suter}},\ }\bibfield  {title} {\bibinfo {title} {Scaling of decoherence in
			wide {NMR} quantum registers},\ }\href
	{https://doi.org/10.1103/PhysRevLett.93.090501} {\bibfield  {journal}
		{\bibinfo  {journal} {Phys. Rev. Lett.}\ }\textbf {\bibinfo {volume} {93}},\
		\bibinfo {pages} {090501} (\bibinfo {year} {2004})}\BibitemShut {NoStop}%
	\bibitem [{\citenamefont {Cho}\ \emph {et~al.}(2005)\citenamefont {Cho},
		\citenamefont {Ladd}, \citenamefont {Baugh}, \citenamefont {Cory},\ and\
		\citenamefont {Ramanathan}}]{Cho2005}%
	\BibitemOpen
	\bibfield  {author} {\bibinfo {author} {\bibfnamefont {H.}~\bibnamefont
			{Cho}}, \bibinfo {author} {\bibfnamefont {T.~D.}\ \bibnamefont {Ladd}},
		\bibinfo {author} {\bibfnamefont {J.}~\bibnamefont {Baugh}}, \bibinfo
		{author} {\bibfnamefont {D.~G.}\ \bibnamefont {Cory}},\ and\ \bibinfo
		{author} {\bibfnamefont {C.}~\bibnamefont {Ramanathan}},\ }\bibfield  {title}
	{\bibinfo {title} {Multispin dynamics of the solid-state {{NMR}} free
			induction decay},\ }\href {https://doi.org/10.1103/PhysRevB.72.054427}
	{\bibfield  {journal} {\bibinfo  {journal} {Phys. Rev. B}\ }\textbf {\bibinfo
			{volume} {72}},\ \bibinfo {pages} {054427} (\bibinfo {year}
		{2005})}\BibitemShut {NoStop}%
	\bibitem [{\citenamefont {van Beek}\ \emph {et~al.}(2005)\citenamefont {van
			Beek}, \citenamefont {Carravetta}, \citenamefont {Antonioli},\ and\
		\citenamefont {Levitt}}]{vanBeek2005}%
	\BibitemOpen
	\bibfield  {author} {\bibinfo {author} {\bibfnamefont {J.~D.}\ \bibnamefont
			{van Beek}}, \bibinfo {author} {\bibfnamefont {M.}~\bibnamefont
			{Carravetta}}, \bibinfo {author} {\bibfnamefont {G.~C.}\ \bibnamefont
			{Antonioli}},\ and\ \bibinfo {author} {\bibfnamefont {M.~H.}\ \bibnamefont
			{Levitt}},\ }\bibfield  {title} {\bibinfo {title} {Spherical tensor analysis
			of nuclear magnetic resonance signals},\ }\href
	{http://scitation.aip.org/content/aip/journal/jcp/122/24/10.1063/1.1943947}
	{\bibfield  {journal} {\bibinfo  {journal} {J. Chem. Phys.}\ }\textbf
		{\bibinfo {volume} {122}},\ \bibinfo {pages} {244510} (\bibinfo {year}
		{2005})}\BibitemShut {NoStop}%
	\bibitem [{\citenamefont {Lovri\ifmmode~\acute{c}\else \'{c}\fi{}}\ \emph
		{et~al.}(2007)\citenamefont {Lovri\ifmmode~\acute{c}\else \'{c}\fi{}},
		\citenamefont {Krojanski},\ and\ \citenamefont {Suter}}]{Lovric2011}%
	\BibitemOpen
	\bibfield  {author} {\bibinfo {author} {\bibfnamefont {M.}~\bibnamefont
			{Lovri\ifmmode~\acute{c}\else \'{c}\fi{}}}, \bibinfo {author} {\bibfnamefont
			{H.~G.}\ \bibnamefont {Krojanski}},\ and\ \bibinfo {author} {\bibfnamefont
			{D.}~\bibnamefont {Suter}},\ }\bibfield  {title} {\bibinfo {title}
		{Decoherence in large quantum registers under variable interaction with the
			environment},\ }\href {https://doi.org/10.1103/PhysRevA.75.042305} {\bibfield
		{journal} {\bibinfo  {journal} {Phys. Rev. A}\ }\textbf {\bibinfo {volume}
			{75}},\ \bibinfo {pages} {042305} (\bibinfo {year} {2007})}\BibitemShut
	{NoStop}%
	\bibitem [{\citenamefont {S{\'a}nchez}\ \emph {et~al.}(2016)\citenamefont
		{S{\'a}nchez}, \citenamefont {Levstein}, \citenamefont {Buljubasich},
		\citenamefont {Pastawski},\ and\ \citenamefont {Chattah}}]{Sanchez2016}%
	\BibitemOpen
	\bibfield  {author} {\bibinfo {author} {\bibfnamefont {C.~M.}\ \bibnamefont
			{S{\'a}nchez}}, \bibinfo {author} {\bibfnamefont {P.~R.}\ \bibnamefont
			{Levstein}}, \bibinfo {author} {\bibfnamefont {L.}~\bibnamefont
			{Buljubasich}}, \bibinfo {author} {\bibfnamefont {H.~M.}\ \bibnamefont
			{Pastawski}},\ and\ \bibinfo {author} {\bibfnamefont {A.~K.}\ \bibnamefont
			{Chattah}},\ }\bibfield  {title} {\bibinfo {title} {Quantum dynamics of
			excitations and decoherence in many-spin systems detected with {Loschmidt}
			echoes: its relation to their spreading through the {Hilbert} space},\ }\href
	{http://rsta.royalsocietypublishing.org/content/roypta/374/2069/20150155.full.pdf}
	{\bibfield  {journal} {\bibinfo  {journal} {Philosophical Transactions of the
				Royal Society A: Mathematical, Physical and Engineering Sciences}\ }\textbf
		{\bibinfo {volume} {374}} (\bibinfo {year} {2016})}\BibitemShut {NoStop}%
	\bibitem [{\citenamefont {G\"arttner}\ \emph {et~al.}(2017)\citenamefont
		{G\"arttner}, \citenamefont {Bohnet}, \citenamefont {Safavi-Naini},
		\citenamefont {Wall}, \citenamefont {Bollinger},\ and\ \citenamefont
		{Rey}}]{Garttner2017}%
	\BibitemOpen
	\bibfield  {author} {\bibinfo {author} {\bibfnamefont {M.}~\bibnamefont
			{G\"arttner}}, \bibinfo {author} {\bibfnamefont {J.~G.}\ \bibnamefont
			{Bohnet}}, \bibinfo {author} {\bibfnamefont {A.}~\bibnamefont
			{Safavi-Naini}}, \bibinfo {author} {\bibfnamefont {M.~L.}\ \bibnamefont
			{Wall}}, \bibinfo {author} {\bibfnamefont {J.~J.}\ \bibnamefont
			{Bollinger}},\ and\ \bibinfo {author} {\bibfnamefont {A.~M.}\ \bibnamefont
			{Rey}},\ }\bibfield  {title} {\bibinfo {title} {Measuring out-of-time-order
			correlations and multiple quantum spectra in a trapped-ion quantum magnet},\
	}\href@noop {} {\bibfield  {journal} {\bibinfo  {journal} {Nat. Phys.}\
		}\textbf {\bibinfo {volume} {13}},\ \bibinfo {pages} {781} (\bibinfo {year}
		{2017})}\BibitemShut {NoStop}%
	\bibitem [{\citenamefont {G\"arttner}\ \emph {et~al.}(2018)\citenamefont
		{G\"arttner}, \citenamefont {Hauke},\ and\ \citenamefont
		{Rey}}]{Garttner2018}%
	\BibitemOpen
	\bibfield  {author} {\bibinfo {author} {\bibfnamefont {M.}~\bibnamefont
			{G\"arttner}}, \bibinfo {author} {\bibfnamefont {P.}~\bibnamefont {Hauke}},\
		and\ \bibinfo {author} {\bibfnamefont {A.~M.}\ \bibnamefont {Rey}},\
	}\bibfield  {title} {\bibinfo {title} {Relating out-of-time-order
			correlations to entanglement via multiple-quantum coherences},\ }\href
	{https://doi.org/10.1103/PhysRevLett.120.040402} {\bibfield  {journal}
		{\bibinfo  {journal} {Phys. Rev. Lett.}\ }\textbf {\bibinfo {volume} {120}},\
		\bibinfo {pages} {040402} (\bibinfo {year} {2018})}\BibitemShut {NoStop}%
	\bibitem [{\citenamefont {de~Gennes}(1958)}]{DeGennes58}%
	\BibitemOpen
	\bibfield  {author} {\bibinfo {author} {\bibfnamefont {P.-G.}\ \bibnamefont
			{de~Gennes}},\ }\bibfield  {title} {\bibinfo {title} {Sur la relaxation
			nucleaire dans les cristaux ioniques},\ }\href
	{https://doi.org/https://doi.org/10.1016/0022-3697(58)90284-1} {\bibfield
		{journal} {\bibinfo  {journal} {Journal of Physics and Chemistry of Solids}\
		}\textbf {\bibinfo {volume} {7}},\ \bibinfo {pages} {345} (\bibinfo {year}
		{1958})}\BibitemShut {NoStop}%
	\bibitem [{\citenamefont {Anderson}(1959)}]{Anderson1959}%
	\BibitemOpen
	\bibfield  {author} {\bibinfo {author} {\bibfnamefont {P.~W.}\ \bibnamefont
			{Anderson}},\ }\bibfield  {title} {\bibinfo {title} {Spectral diffusion,
			phonons, and paramagnetic spin-lattice relaxation},\ }\href
	{https://doi.org/10.1103/PhysRev.114.1002} {\bibfield  {journal} {\bibinfo
			{journal} {Phys. Rev.}\ }\textbf {\bibinfo {volume} {114}},\ \bibinfo {pages}
		{1002} (\bibinfo {year} {1959})}\BibitemShut {NoStop}%
	\bibitem [{\citenamefont {{Gaudin, M.}}(1976)}]{Gaudin76}%
	\BibitemOpen
	\bibfield  {author} {\bibinfo {author} {\bibnamefont {{Gaudin, M.}}},\
	}\bibfield  {title} {\bibinfo {title} {Diagonalisation d'une classe
			d'hamiltoniens de spin},\ }\href
	{https://doi.org/10.1051/jphys:0197600370100108700} {\bibfield  {journal}
		{\bibinfo  {journal} {J. Phys. France}\ }\textbf {\bibinfo {volume} {37}},\
		\bibinfo {pages} {1087} (\bibinfo {year} {1976})}\BibitemShut {NoStop}%
	\bibitem [{\citenamefont {Prokof'ev}\ and\ \citenamefont
		{Stamp}(2000)}]{Stamp2000}%
	\BibitemOpen
	\bibfield  {author} {\bibinfo {author} {\bibfnamefont {N.~V.}\ \bibnamefont
			{Prokof'ev}}\ and\ \bibinfo {author} {\bibfnamefont {P.~C.~E.}\ \bibnamefont
			{Stamp}},\ }\bibfield  {title} {\bibinfo {title} {Theory of the spin bath},\
	}\href {http://stacks.iop.org/0034-4885/63/i=4/a=204} {\bibfield  {journal}
		{\bibinfo  {journal} {Reports on Progress in Physics}\ }\textbf {\bibinfo
			{volume} {63}},\ \bibinfo {pages} {669} (\bibinfo {year} {2000})}\BibitemShut
	{NoStop}%
	\bibitem [{\citenamefont {Rhim}\ \emph {et~al.}(1973)\citenamefont {Rhim},
		\citenamefont {Elleman},\ and\ \citenamefont {Vaughan}}]{Rhim73_mrev8}%
	\BibitemOpen
	\bibfield  {author} {\bibinfo {author} {\bibfnamefont {W.}~\bibnamefont
			{Rhim}}, \bibinfo {author} {\bibfnamefont {D.~D.}\ \bibnamefont {Elleman}},\
		and\ \bibinfo {author} {\bibfnamefont {R.~W.}\ \bibnamefont {Vaughan}},\
	}\bibfield  {title} {\bibinfo {title} {Enhanced resolution for solid state
			{NMR}},\ }\href {http://aip.scitation.org/doi/abs/10.1063/1.1679423}
	{\bibfield  {journal} {\bibinfo  {journal} {J. Chem. Phys.}\ }\textbf
		{\bibinfo {volume} {58}},\ \bibinfo {pages} {1772} (\bibinfo {year}
		{1973})}\BibitemShut {NoStop}%
	\bibitem [{\citenamefont {Rhim}\ \emph {et~al.}(1974)\citenamefont {Rhim},
		\citenamefont {Elleman}, \citenamefont {Schreiber},\ and\ \citenamefont
		{Vaughan}}]{Rhim74_tuneupcycles}%
	\BibitemOpen
	\bibfield  {author} {\bibinfo {author} {\bibfnamefont {W.}~\bibnamefont
			{Rhim}}, \bibinfo {author} {\bibfnamefont {D.~D.}\ \bibnamefont {Elleman}},
		\bibinfo {author} {\bibfnamefont {L.~B.}\ \bibnamefont {Schreiber}},\ and\
		\bibinfo {author} {\bibfnamefont {R.~W.}\ \bibnamefont {Vaughan}},\
	}\bibfield  {title} {\bibinfo {title} {Analysis of multiple pulse {NMR} in
			solids. {II}},\ }\href
	{http://scitation.aip.org/content/aip/journal/jcp/60/11/10.1063/1.1680944}
	{\bibfield  {journal} {\bibinfo  {journal} {The Journal of Chemical Physics}\
		}\textbf {\bibinfo {volume} {60}},\ \bibinfo {pages} {4595} (\bibinfo {year}
		{1974})}\BibitemShut {NoStop}%
	\bibitem [{\citenamefont {Haeberlen}(1976)}]{HaeberlenBook}%
	\BibitemOpen
	\bibfield  {author} {\bibinfo {author} {\bibfnamefont {U.}~\bibnamefont
			{Haeberlen}},\ }\href@noop {} {\emph {\bibinfo {title} {High Resolution
				{{{NMR}}} in Solids: Selective Averaging}}}\ (\bibinfo  {publisher} {Academic
		Press},\ \bibinfo {address} {New York},\ \bibinfo {year} {1976})\BibitemShut
	{NoStop}%
	\bibitem [{\citenamefont {Gerstein}\ and\ \citenamefont
		{Dybowski}(1985)}]{DybowskiBook}%
	\BibitemOpen
	\bibfield  {author} {\bibinfo {author} {\bibfnamefont {B.~C.}\ \bibnamefont
			{Gerstein}}\ and\ \bibinfo {author} {\bibfnamefont {C.~R.}\ \bibnamefont
			{Dybowski}},\ }\href@noop {} {\emph {\bibinfo {title} {Transient techniques
				in {{NMR}} of solids}}}\ (\bibinfo  {publisher} {Academic Press, Inc.},\
	\bibinfo {year} {1985})\BibitemShut {NoStop}%
	\bibitem [{\citenamefont {Li}\ \emph {et~al.}(2017)\citenamefont {Li},
		\citenamefont {Fan}, \citenamefont {Wang}, \citenamefont {Ye}, \citenamefont
		{Zeng}, \citenamefont {Zhai}, \citenamefont {Peng},\ and\ \citenamefont
		{Du}}]{Li2017}%
	\BibitemOpen
	\bibfield  {author} {\bibinfo {author} {\bibfnamefont {J.}~\bibnamefont
			{Li}}, \bibinfo {author} {\bibfnamefont {R.}~\bibnamefont {Fan}}, \bibinfo
		{author} {\bibfnamefont {H.}~\bibnamefont {Wang}}, \bibinfo {author}
		{\bibfnamefont {B.}~\bibnamefont {Ye}}, \bibinfo {author} {\bibfnamefont
			{B.}~\bibnamefont {Zeng}}, \bibinfo {author} {\bibfnamefont {H.}~\bibnamefont
			{Zhai}}, \bibinfo {author} {\bibfnamefont {X.}~\bibnamefont {Peng}},\ and\
		\bibinfo {author} {\bibfnamefont {J.}~\bibnamefont {Du}},\ }\bibfield
	{title} {\bibinfo {title} {Measuring out-of-time-order correlators on a
			nuclear magnetic resonance quantum simulator},\ }\href
	{https://doi.org/10.1103/PhysRevX.7.031011} {\bibfield  {journal} {\bibinfo
			{journal} {Phys. Rev. X}\ }\textbf {\bibinfo {volume} {7}},\ \bibinfo {pages}
		{031011} (\bibinfo {year} {2017})}\BibitemShut {NoStop}%
	\bibitem [{\citenamefont {Wei}\ \emph {et~al.}(2018)\citenamefont {Wei},
		\citenamefont {Ramanathan},\ and\ \citenamefont {Cappellaro}}]{Wei2018}%
	\BibitemOpen
	\bibfield  {author} {\bibinfo {author} {\bibfnamefont {K.~X.}\ \bibnamefont
			{Wei}}, \bibinfo {author} {\bibfnamefont {C.}~\bibnamefont {Ramanathan}},\
		and\ \bibinfo {author} {\bibfnamefont {P.}~\bibnamefont {Cappellaro}},\
	}\bibfield  {title} {\bibinfo {title} {Exploring localization in nuclear spin
			chains},\ }\href {https://doi.org/10.1103/PhysRevLett.120.070501} {\bibfield
		{journal} {\bibinfo  {journal} {Phys. Rev. Lett.}\ }\textbf {\bibinfo
			{volume} {120}},\ \bibinfo {pages} {070501} (\bibinfo {year}
		{2018})}\BibitemShut {NoStop}%
	\bibitem [{\citenamefont {Kitaev}()}]{Kitaev2015}%
	\BibitemOpen
	\bibfield  {author} {\bibinfo {author} {\bibfnamefont {A.~Y.}\ \bibnamefont
			{Kitaev}},\ }\bibfield  {title} {\bibinfo {title} {A simple model of quantum
			holography},\ }\bibinfo {note}
	{http://online.kitp.ucsb.edu/online/entangled15/kitaev/ \&
		http://online.kitp.ucsb.edu/online/entangled15/kitaev2/}\BibitemShut
	{NoStop}%
	\bibitem [{\citenamefont {Maldacena}\ \emph {et~al.}(2016)\citenamefont
		{Maldacena}, \citenamefont {Shenker},\ and\ \citenamefont
		{Stanford}}]{Maldacena2016}%
	\BibitemOpen
	\bibfield  {author} {\bibinfo {author} {\bibfnamefont {J.}~\bibnamefont
			{Maldacena}}, \bibinfo {author} {\bibfnamefont {S.~H.}\ \bibnamefont
			{Shenker}},\ and\ \bibinfo {author} {\bibfnamefont {D.}~\bibnamefont
			{Stanford}},\ }\bibfield  {title} {\bibinfo {title} {A bound on chaos},\
	}\href {https://doi.org/10.1007/JHEP08(2016)106} {\bibfield  {journal}
		{\bibinfo  {journal} {J. High Energ. Phys.}\ }\textbf {\bibinfo {volume}
			{2016}}\bibinfo  {number} { (8)},\ \bibinfo {pages} {106}}\BibitemShut
	{NoStop}%
	\bibitem [{\citenamefont {Rozenbaum}\ \emph {et~al.}(2017)\citenamefont
		{Rozenbaum}, \citenamefont {Ganeshan},\ and\ \citenamefont
		{Galitski}}]{Rozenbaum2017}%
	\BibitemOpen
	\bibfield  {number} {  }\bibfield  {author} {\bibinfo {author} {\bibfnamefont
			{E.~B.}\ \bibnamefont {Rozenbaum}}, \bibinfo {author} {\bibfnamefont
			{S.}~\bibnamefont {Ganeshan}},\ and\ \bibinfo {author} {\bibfnamefont
			{V.}~\bibnamefont {Galitski}},\ }\bibfield  {title} {\bibinfo {title}
		{{Lyapunov} exponent and out-of-time-ordered correlator's growth rate in a
			chaotic system},\ }\href {https://doi.org/10.1103/PhysRevLett.118.086801}
	{\bibfield  {journal} {\bibinfo  {journal} {Phys. Rev. Lett.}\ }\textbf
		{\bibinfo {volume} {118}},\ \bibinfo {pages} {086801} (\bibinfo {year}
		{2017})}\BibitemShut {NoStop}%
	\bibitem [{\citenamefont {{Rozenbaum}}\ \emph {et~al.}(2018)\citenamefont
		{{Rozenbaum}}, \citenamefont {{Ganeshan}},\ and\ \citenamefont
		{{Galitski}}}]{Rozenbaum18}%
	\BibitemOpen
	\bibfield  {author} {\bibinfo {author} {\bibfnamefont {E.~B.}\ \bibnamefont
			{{Rozenbaum}}}, \bibinfo {author} {\bibfnamefont {S.}~\bibnamefont
			{{Ganeshan}}},\ and\ \bibinfo {author} {\bibfnamefont {V.}~\bibnamefont
			{{Galitski}}},\ }\bibfield  {title} {\bibinfo {title} {Universal level
			statistics of the out-of-time-ordered operator},\ }\href@noop {} {\
		(\bibinfo {year} {2018})},\ \Eprint {https://arxiv.org/abs/1801.10591}
	{arXiv:1801.10591} \BibitemShut {NoStop}%
	\bibitem [{\citenamefont {Ch\'avez-Carlos}\ \emph {et~al.}(2019)\citenamefont
		{Ch\'avez-Carlos}, \citenamefont {L\'opez-del Carpio}, \citenamefont
		{Bastarrachea-Magnani}, \citenamefont {Str\'ansk\'y}, \citenamefont
		{Lerma-Hern\'andez}, \citenamefont {Santos},\ and\ \citenamefont
		{Hirsch}}]{Chavez-Carlos:2018}%
	\BibitemOpen
	\bibfield  {author} {\bibinfo {author} {\bibfnamefont {J.}~\bibnamefont
			{Ch\'avez-Carlos}}, \bibinfo {author} {\bibfnamefont {B.}~\bibnamefont
			{L\'opez-del Carpio}}, \bibinfo {author} {\bibfnamefont {M.~A.}\ \bibnamefont
			{Bastarrachea-Magnani}}, \bibinfo {author} {\bibfnamefont {P.}~\bibnamefont
			{Str\'ansk\'y}}, \bibinfo {author} {\bibfnamefont {S.}~\bibnamefont
			{Lerma-Hern\'andez}}, \bibinfo {author} {\bibfnamefont {L.~F.}\ \bibnamefont
			{Santos}},\ and\ \bibinfo {author} {\bibfnamefont {J.~G.}\ \bibnamefont
			{Hirsch}},\ }\bibfield  {title} {\bibinfo {title} {Quantum and classical
			{Lyapunov} exponents in atom-field interaction systems},\ }\href
	{https://doi.org/10.1103/PhysRevLett.122.024101} {\bibfield  {journal}
		{\bibinfo  {journal} {Phys. Rev. Lett.}\ }\textbf {\bibinfo {volume} {122}},\
		\bibinfo {pages} {024101} (\bibinfo {year} {2019})}\BibitemShut {NoStop}%
	\bibitem [{\citenamefont {Usaj}\ \emph {et~al.}(1998)\citenamefont {Usaj},
		\citenamefont {Pastawski},\ and\ \citenamefont
		{Levstein}}]{Usaj-Levstein1998}%
	\BibitemOpen
	\bibfield  {author} {\bibinfo {author} {\bibfnamefont {G.}~\bibnamefont
			{Usaj}}, \bibinfo {author} {\bibfnamefont {H.~M.}\ \bibnamefont
			{Pastawski}},\ and\ \bibinfo {author} {\bibfnamefont {P.~R.}\ \bibnamefont
			{Levstein}},\ }\bibfield  {title} {\bibinfo {title} {Gaussian to exponential
			crossover in the attenuation of polarization echoes in {NMR}},\ }\bibfield
	{booktitle} {\emph {\bibinfo {booktitle} {Molecular Physics}},\ }\href
	{https://doi.org/10.1080/00268979809483253} {\bibfield  {journal} {\bibinfo
			{journal} {Molecular Physics}\ }\textbf {\bibinfo {volume} {95}},\ \bibinfo
		{pages} {1229} (\bibinfo {year} {1998})}\BibitemShut {NoStop}%
	\bibitem [{\citenamefont {Levstein}\ \emph {et~al.}(1998)\citenamefont
		{Levstein}, \citenamefont {Usaj},\ and\ \citenamefont
		{Pastawski}}]{Levstein-Pastawski1998}%
	\BibitemOpen
	\bibfield  {author} {\bibinfo {author} {\bibfnamefont {P.~R.}\ \bibnamefont
			{Levstein}}, \bibinfo {author} {\bibfnamefont {G.}~\bibnamefont {Usaj}},\
		and\ \bibinfo {author} {\bibfnamefont {H.~M.}\ \bibnamefont {Pastawski}},\
	}\bibfield  {title} {\bibinfo {title} {Attenuation of polarization echoes in
			nuclear magnetic resonance: A study of the emergence of dynamical
			irreversibility in many-body quantum systems},\ }\bibfield  {booktitle}
	{\emph {\bibinfo {booktitle} {The Journal of Chemical Physics}},\ }\href
	{https://doi.org/10.1063/1.475664} {\bibfield  {journal} {\bibinfo  {journal}
			{The Journal of Chemical Physics}\ }\textbf {\bibinfo {volume} {108}},\
		\bibinfo {pages} {2718} (\bibinfo {year} {1998})}\BibitemShut {NoStop}%
	\bibitem [{\citenamefont {Lu}\ \emph {et~al.}(2010)\citenamefont {Lu},
		\citenamefont {Wang},\ and\ \citenamefont {Sun}}]{Lu2010}%
	\BibitemOpen
	\bibfield  {author} {\bibinfo {author} {\bibfnamefont {X.-M.}\ \bibnamefont
			{Lu}}, \bibinfo {author} {\bibfnamefont {X.}~\bibnamefont {Wang}},\ and\
		\bibinfo {author} {\bibfnamefont {C.~P.}\ \bibnamefont {Sun}},\ }\bibfield
	{title} {\bibinfo {title} {Quantum {Fisher} information flow and
			non-{Markovian} processes of open systems},\ }\href
	{https://doi.org/10.1103/PhysRevA.82.042103} {\bibfield  {journal} {\bibinfo
			{journal} {Phys. Rev. A}\ }\textbf {\bibinfo {volume} {82}},\ \bibinfo
		{pages} {042103} (\bibinfo {year} {2010})}\BibitemShut {NoStop}%
	\bibitem [{\citenamefont {Abragam}(1961)}]{Abragam}%
	\BibitemOpen
	\bibfield  {author} {\bibinfo {author} {\bibfnamefont {A.}~\bibnamefont
			{Abragam}},\ }\href@noop {} {\emph {\bibinfo {title} {The Principles of
				Nuclear Magnetism}}}\ (\bibinfo  {publisher} {Oxford University Press},\
	\bibinfo {year} {1961})\BibitemShut {NoStop}%
	\bibitem [{\citenamefont {Cho}\ \emph {et~al.}(2006)\citenamefont {Cho},
		\citenamefont {Cappellaro}, \citenamefont {Cory},\ and\ \citenamefont
		{Ramanathan}}]{Cho2006}%
	\BibitemOpen
	\bibfield  {author} {\bibinfo {author} {\bibfnamefont {H.}~\bibnamefont
			{Cho}}, \bibinfo {author} {\bibfnamefont {P.}~\bibnamefont {Cappellaro}},
		\bibinfo {author} {\bibfnamefont {D.~G.}\ \bibnamefont {Cory}},\ and\
		\bibinfo {author} {\bibfnamefont {C.}~\bibnamefont {Ramanathan}},\ }\bibfield
	{title} {\bibinfo {title} {Decay of highly correlated spin states in a
			dipolar-coupled solid: {NMR} study of $\mathrm{CaF}_{2}$},\ }\href
	{https://doi.org/10.1103/PhysRevB.74.224434} {\bibfield  {journal} {\bibinfo
			{journal} {Phys. Rev. B}\ }\textbf {\bibinfo {volume} {74}},\ \bibinfo
		{pages} {224434} (\bibinfo {year} {2006})}\BibitemShut {NoStop}%
	\bibitem [{\citenamefont {Peres}(1996)}]{Peres1996}%
	\BibitemOpen
	\bibfield  {author} {\bibinfo {author} {\bibfnamefont {A.}~\bibnamefont
			{Peres}},\ }\bibfield  {title} {\bibinfo {title} {Chaotic evolution in
			quantum mechanics},\ }\href {https://doi.org/10.1103/PhysRevE.53.4524}
	{\bibfield  {journal} {\bibinfo  {journal} {Phys. Rev. E}\ }\textbf {\bibinfo
			{volume} {53}},\ \bibinfo {pages} {4524} (\bibinfo {year}
		{1996})}\BibitemShut {NoStop}%
	\bibitem [{\citenamefont {Cucchietti}\ \emph {et~al.}(2002)\citenamefont
		{Cucchietti}, \citenamefont {Lewenkopf}, \citenamefont {Mucciolo},
		\citenamefont {Pastawski},\ and\ \citenamefont {Vallejos}}]{Cucchietti2002}%
	\BibitemOpen
	\bibfield  {author} {\bibinfo {author} {\bibfnamefont {F.~M.}\ \bibnamefont
			{Cucchietti}}, \bibinfo {author} {\bibfnamefont {C.~H.}\ \bibnamefont
			{Lewenkopf}}, \bibinfo {author} {\bibfnamefont {E.~R.}\ \bibnamefont
			{Mucciolo}}, \bibinfo {author} {\bibfnamefont {H.~M.}\ \bibnamefont
			{Pastawski}},\ and\ \bibinfo {author} {\bibfnamefont {R.~O.}\ \bibnamefont
			{Vallejos}},\ }\bibfield  {title} {\bibinfo {title} {Measuring the {Lyapunov}
			exponent using quantum mechanics},\ }\href@noop {} {\bibfield  {journal}
		{\bibinfo  {journal} {Phys. Rev. E}\ }\textbf {\bibinfo {volume} {65}},\
		\bibinfo {pages} {046209} (\bibinfo {year} {2002})}\BibitemShut {NoStop}%
	\bibitem [{\citenamefont {Emerson}\ \emph {et~al.}(2002)\citenamefont
		{Emerson}, \citenamefont {Weinstein}, \citenamefont {Lloyd},\ and\
		\citenamefont {Cory}}]{Emerson2002}%
	\BibitemOpen
	\bibfield  {author} {\bibinfo {author} {\bibfnamefont {J.}~\bibnamefont
			{Emerson}}, \bibinfo {author} {\bibfnamefont {Y.~S.}\ \bibnamefont
			{Weinstein}}, \bibinfo {author} {\bibfnamefont {S.}~\bibnamefont {Lloyd}},\
		and\ \bibinfo {author} {\bibfnamefont {D.~G.}\ \bibnamefont {Cory}},\
	}\bibfield  {title} {\bibinfo {title} {Fidelity decay as an efficient
			indicator of quantum chaos},\ }\href@noop {} {\bibfield  {journal} {\bibinfo
			{journal} {Phys. Rev. Lett.}\ }\textbf {\bibinfo {volume} {89}},\ \bibinfo
		{pages} {284102} (\bibinfo {year} {2002})}\BibitemShut {NoStop}%
	\bibitem [{\citenamefont {Gorin}\ \emph {et~al.}(2006)\citenamefont {Gorin},
		\citenamefont {Prosen}, \citenamefont {Seligman},\ and\ \citenamefont
		{\ifmmode \check{Z}\else \v{Z}\fi{}nidari\ifmmode~\check{c}\else
			\v{c}\fi{}}}]{Gorin2006}%
	\BibitemOpen
	\bibfield  {author} {\bibinfo {author} {\bibfnamefont {T.}~\bibnamefont
			{Gorin}}, \bibinfo {author} {\bibfnamefont {T.}~\bibnamefont {Prosen}},
		\bibinfo {author} {\bibfnamefont {T.~H.}\ \bibnamefont {Seligman}},\ and\
		\bibinfo {author} {\bibfnamefont {M.}~\bibnamefont {\ifmmode \check{Z}\else
				\v{Z}\fi{}nidari\ifmmode~\check{c}\else \v{c}\fi{}}},\ }\bibfield  {title}
	{\bibinfo {title} {Dynamics of {Loschmidt} echoes and fidelity decay},\
	}\href@noop {} {\bibfield  {journal} {\bibinfo  {journal} {Phys. Rep.}\
		}\textbf {\bibinfo {volume} {435}},\ \bibinfo {pages} {33 } (\bibinfo {year}
		{2006})}\BibitemShut {NoStop}%
	\bibitem [{\citenamefont {Elsayed}\ and\ \citenamefont
		{Fine}(2015)}]{Elsayed2015}%
	\BibitemOpen
	\bibfield  {author} {\bibinfo {author} {\bibfnamefont {T.~A.}\ \bibnamefont
			{Elsayed}}\ and\ \bibinfo {author} {\bibfnamefont {B.~V.}\ \bibnamefont
			{Fine}},\ }\bibfield  {title} {\bibinfo {title} {Sensitivity to small
			perturbations in systems of large quantum spins},\ }\href
	{http://stacks.iop.org/1402-4896/2015/i=T165/a=014011} {\bibfield  {journal}
		{\bibinfo  {journal} {Phys. Scr.}\ }\textbf {\bibinfo {volume} {2015}},\
		\bibinfo {pages} {014011} (\bibinfo {year} {2015})}\BibitemShut {NoStop}%
	\bibitem [{\citenamefont {Borgonovi}\ \emph {et~al.}(2019)\citenamefont
		{Borgonovi}, \citenamefont {Izrailev},\ and\ \citenamefont
		{Santos}}]{Borgonovi2019}%
	\BibitemOpen
	\bibfield  {author} {\bibinfo {author} {\bibfnamefont {F.}~\bibnamefont
			{Borgonovi}}, \bibinfo {author} {\bibfnamefont {F.~M.}\ \bibnamefont
			{Izrailev}},\ and\ \bibinfo {author} {\bibfnamefont {L.~F.}\ \bibnamefont
			{Santos}},\ }\bibfield  {title} {\bibinfo {title} {Exponentially fast
			dynamics of chaotic many-body systems},\ }\href
	{https://doi.org/10.1103/PhysRevE.99.010101} {\bibfield  {journal} {\bibinfo
			{journal} {Phys. Rev. E}\ }\textbf {\bibinfo {volume} {99}},\ \bibinfo
		{pages} {010101} (\bibinfo {year} {2019})}\BibitemShut {NoStop}%
			\bibitem [{\citenamefont {Chirikov}\ \emph {et~al.}(1981)\citenamefont
		{Chirikov}, \citenamefont {Izrailev},\ and\ \citenamefont
		{Shepelyansky}}]{Chirikov1981}%
	\BibitemOpen
	\bibfield  {author} {\bibinfo {author} {\bibfnamefont {B.~V.}\ \bibnamefont
			{Chirikov}}, \bibinfo {author} {\bibfnamefont {F.~M.}\ \bibnamefont
			{Izrailev}},\ and\ \bibinfo {author} {\bibfnamefont {D.~L.}\ \bibnamefont
			{Shepelyansky}},\ }\bibfield  {title} {\bibinfo {title} {Dynamical
			stochasticity in classical and quantum mechanics},\ }\href@noop {} {\bibfield
		{journal} {\bibinfo  {journal} {Sov. Sci. Rev.}\ }\textbf {\bibinfo {volume}
			{C 2}},\ \bibinfo {pages} {209} (\bibinfo {year} {1981})}\BibitemShut
	{NoStop}%
	\bibitem [{\citenamefont {Shepelyansky}(1983)}]{SHEPELYANSKY1983}%
	\BibitemOpen
	\bibfield  {author} {\bibinfo {author} {\bibfnamefont {D.}~\bibnamefont
			{Shepelyansky}},\ }\bibfield  {title} {\bibinfo {title} {Some statistical
			properties of simple classically stochastic quantum systems},\ }\href
	{https://doi.org/https://doi.org/10.1016/0167-2789(83)90318-4} {\bibfield
		{journal} {\bibinfo  {journal} {Physica D: Nonlinear Phenomena}\ }\textbf
		{\bibinfo {volume} {8}},\ \bibinfo {pages} {208 } (\bibinfo {year}
		{1983})}\BibitemShut {NoStop}%
	\bibitem [{\citenamefont {Flambaum}\ and\ \citenamefont
		{Izrailev}(2001)}]{Flambaum2001}%
	\BibitemOpen
	\bibfield  {author} {\bibinfo {author} {\bibfnamefont {V.~V.}\ \bibnamefont
			{Flambaum}}\ and\ \bibinfo {author} {\bibfnamefont {F.~M.}\ \bibnamefont
			{Izrailev}},\ }\bibfield  {title} {\bibinfo {title} {Entropy production and
			wave packet dynamics in the {Fock} space of closed chaotic many-body
			systems},\ }\href {https://doi.org/10.1103/PhysRevE.64.036220} {\bibfield
		{journal} {\bibinfo  {journal} {Phys. Rev. E}\ }\textbf {\bibinfo {volume}
			{64}},\ \bibinfo {pages} {036220} (\bibinfo {year} {2001})}\BibitemShut
	{NoStop}%
	\bibitem [{\citenamefont {Garc\'{\i}a-Mata}\ \emph {et~al.}(2018)\citenamefont
		{Garc\'{\i}a-Mata}, \citenamefont {Saraceno}, \citenamefont {Jalabert},
		\citenamefont {Roncaglia},\ and\ \citenamefont
		{Wisniacki}}]{Garcia-Mata2018}%
	\BibitemOpen
	\bibfield  {author} {\bibinfo {author} {\bibfnamefont {I.}~\bibnamefont
			{Garc\'{\i}a-Mata}}, \bibinfo {author} {\bibfnamefont {M.}~\bibnamefont
			{Saraceno}}, \bibinfo {author} {\bibfnamefont {R.~A.}\ \bibnamefont
			{Jalabert}}, \bibinfo {author} {\bibfnamefont {A.~J.}\ \bibnamefont
			{Roncaglia}},\ and\ \bibinfo {author} {\bibfnamefont {D.~A.}\ \bibnamefont
			{Wisniacki}},\ }\bibfield  {title} {\bibinfo {title} {Chaos signatures in the
			short and long time behavior of the out-of-time ordered correlator},\ }\href
	{https://doi.org/10.1103/PhysRevLett.121.210601} {\bibfield  {journal}
		{\bibinfo  {journal} {Phys. Rev. Lett.}\ }\textbf {\bibinfo {volume} {121}},\
		\bibinfo {pages} {210601} (\bibinfo {year} {2018})}\BibitemShut {NoStop}%
	\bibitem [{\citenamefont {Luitz}\ and\ \citenamefont
		{Bar~Lev}(2017)}]{Luitz2017}%
	\BibitemOpen
	\bibfield  {author} {\bibinfo {author} {\bibfnamefont {D.~J.}\ \bibnamefont
			{Luitz}}\ and\ \bibinfo {author} {\bibfnamefont {Y.}~\bibnamefont
			{Bar~Lev}},\ }\bibfield  {title} {\bibinfo {title} {Information propagation
			in isolated quantum systems},\ }\href
	{https://doi.org/10.1103/PhysRevB.96.020406} {\bibfield  {journal} {\bibinfo
			{journal} {Phys. Rev. B}\ }\textbf {\bibinfo {volume} {96}},\ \bibinfo
		{pages} {020406} (\bibinfo {year} {2017})}\BibitemShut {NoStop}%
	\bibitem [{\citenamefont {Niknam}(2018)}]{Mohamad_PhDthesis}%
	\BibitemOpen
	\bibfield  {author} {\bibinfo {author} {\bibfnamefont {M.}~\bibnamefont
			{Niknam}},\ }\emph {\bibinfo {title} {Dynamics of Quantum Information of the
			Central Spin Problem}},\ \href {http://hdl.handle.net/10012/12870} {Ph.D.
		thesis},\ \bibinfo  {school} {University of Waterloo}, \bibinfo {address}
	{http://hdl.handle.net/10012/12870} (\bibinfo {year} {2018})\BibitemShut
	{NoStop}%
	\bibitem [{\citenamefont {Guhr}\ \emph {et~al.}(1998)\citenamefont {Guhr},
		\citenamefont {Mueller-Gr\"oeling},\ and\ \citenamefont
		{Weidenm\"uller}}]{Guhr1998}%
	\BibitemOpen
	\bibfield  {author} {\bibinfo {author} {\bibfnamefont {T.}~\bibnamefont
			{Guhr}}, \bibinfo {author} {\bibfnamefont {A.}~\bibnamefont
			{Mueller-Gr\"oeling}},\ and\ \bibinfo {author} {\bibfnamefont {H.~A.}\
			\bibnamefont {Weidenm\"uller}},\ }\bibfield  {title} {\bibinfo {title}
		{Random matrix theories in quantum physics: Common concepts},\ }\href@noop {}
	{\bibfield  {journal} {\bibinfo  {journal} {Phys. Rep.}\ }\textbf {\bibinfo
			{volume} {299}},\ \bibinfo {pages} {189} (\bibinfo {year}
		{1998})}\BibitemShut {NoStop}%
	\bibitem [{\citenamefont {Borgonovi}\ \emph {et~al.}(2016)\citenamefont
		{Borgonovi}, \citenamefont {Izrailev}, \citenamefont {Santos},\ and\
		\citenamefont {Zelevinsky}}]{Borgonovi2016}%
	\BibitemOpen
	\bibfield  {author} {\bibinfo {author} {\bibfnamefont {F.}~\bibnamefont
			{Borgonovi}}, \bibinfo {author} {\bibfnamefont {F.~M.}\ \bibnamefont
			{Izrailev}}, \bibinfo {author} {\bibfnamefont {L.~F.}\ \bibnamefont
			{Santos}},\ and\ \bibinfo {author} {\bibfnamefont {V.~G.}\ \bibnamefont
			{Zelevinsky}},\ }\bibfield  {title} {\bibinfo {title} {Quantum chaos and
			thermalization in isolated systems of interacting particles},\ }\href
	{https://doi.org/10.1016/j.physrep.2016.02.005} {\bibfield  {journal}
		{\bibinfo  {journal} {Phys. Rep.}\ }\textbf {\bibinfo {volume} {626}},\
		\bibinfo {pages} {1} (\bibinfo {year} {2016})}\BibitemShut {NoStop}%
\end{thebibliography}

%


\end{document}